\documentclass[11pt]{article}
\usepackage[dvipdfm]{graphicx}
\usepackage{geometry}
\usepackage{authblk}
\usepackage{url}
\usepackage{multirow}%
\usepackage{amsmath,amssymb,amsfonts}%
\usepackage{amsthm}%
\usepackage{mathrsfs}%
\usepackage[title]{appendix}%
\usepackage{xcolor}%
\usepackage{textcomp}%
\usepackage{manyfoot}%
\usepackage{booktabs}%
\usepackage{authblk}
\usepackage{longtable}
\usepackage{caption}
\usepackage{color}
\usepackage{algorithm}
\usepackage[noend]{algorithmic}
\usepackage{listings}%
\usepackage{afterpage,array,rotating}

\newcolumntype{A}{>{\centering\arraybackslash}p{2.1cm}}
\newcolumntype{B}{>{\centering\arraybackslash}p{1.5cm}}
\newcolumntype{C}{>{\centering\arraybackslash}p{1.4cm}}
\newcolumntype{D}{>{\centering\arraybackslash}p{3.5cm}}
\newcolumntype{E}{>{\centering\arraybackslash}p{2.5cm}}

\title{Systemic Gendered Citation Imbalance in Computer Science: Evidence from Conferences and Journals}

\author[1]{Kazuki Nakajima}
\author[2]{Yuya Sasaki}
\author[3]{Sohei Tokuno}
\author[4]{George Fletcher}

\affil[1]{Tokyo Metropolitan University}
\affil[2]{The University of Osaka}
\affil[3]{Nara Institute of Science and Technology}
\affil[4]{Eindhoven University of Technology}

\begin{document}
\date{}
\maketitle

\begin{abstract}
Gender imbalance persists across science, technology, engineering, and mathematics (STEM) fields, including computer science, where it appears in researcher demographics, productivity, recognition, hiring, and career progression.  
Given computer science's rapid expansion and global influence, addressing this imbalance is essential for broadening participation and fueling innovation.  
Although journal-oriented disciplines exhibit consistent gender imbalances in citation practices, it remains unclear whether similar patterns arise in the conference-centric culture of computer science.  
Here, we systematically investigate gender imbalance in citations of conference and journal papers in computer science. 
We find that papers for which a woman is listed as either first or last author receive fewer citations than expected, partly because of homophilic citation tendencies (i.e., authors tend to cite papers that share specific attributes).  
This imbalance is especially pronounced for conference papers--particularly those published at top-tier venues--relative to journals.  
Moreover, we find that the prominence of the first or last author and the structure of their local co-authorship networks are potential drivers of these imbalances.  
By exploring how conference-centric publishing practices can amplify systemic imbalances in computer science, our study offers insights that may inform efforts to foster more equitable representation in academia.
\end{abstract}


\section{Introduction} 

Gender imbalance remains a pervasive issue across academia worldwide. 
Although the proportion of female authors has been steadily increasing, women remain underrepresented in senior authorship positions and leadership roles across multiple disciplines and regions, compared to the gender ratio at undergraduate and graduate levels \cite{lariviere2013, holman2018, sanchez_jimenez2024}.
In addition, systemic disparities persist in research productivity \cite{xie1998, besselaar2016, huang2020, abramo2021}, academic awardees \cite{lincoln2012, meho2021}, faculty hiring \cite{corinne2012, sheltzer2014}, access to funding \cite{bornmann2007, witteman2019}, and citations received by authors \cite{ioannidis2023}, all of which collectively disadvantage female researchers throughout their careers. 
Observed across various disciplines, these imbalances reflect a broader pattern of inequity embedded within the academic ecosystem.
Addressing such systemic imbalances is essential for achieving equity and sustainability in academia and promoting greater scientific diversity and innovation \cite{llorens2021, sugimoto2023}.

Citations form the basis for many measurements and indicators in ``science of science'' studies, informing assessments of the impact of publications, researchers, institutions, and countries, and key decisions in academic evaluation \cite{fortunato2018, milojevic2025}.
While citations offer valuable insights into the structure and development of scientific knowledge, citation statistics are often shaped by a range of social, structural, and epistemic factors that introduce systematic imbalances.
For example, average citation counts vary substantially across disciplines \cite{radicchi2008}, and many journals receive a disproportionate share of their citations from a small set of publication venues \cite{kojaku2021}.
Similarly, research-active countries tend to receive more citations than expected given the subject areas in which they publish \cite{gomez2022}.
In the context of gender, recent studies have identified consistent under-citation of women-authored papers, partly due to homophily, skewed citation practices, and self-citation patterns favoring male authors \cite{king2017, dworkin2020, teich2022}.
Such imbalances distort the production and flow of scientific knowledge and undermine the neutrality of citations as indicators of scholarly value.

Computer science stands out globally for its vast and expanding student and researcher population, extensive research output, and significant investment. 
First, the pipeline of human resources in science, technology, engineering, and mathematics (STEM)--including computer science--is immense and growing. 
In 2020 alone, China awarded 3.6 million new STEM degrees, India 2.6 million, and the United States 820,000, with countries such as Brazil and Mexico outpacing Iran and Japan \cite{oliss2023}. 
This growth is often driven by computer science programs; for example, North African nations like Tunisia and Algeria report high STEM graduation rates owing to robust computing and engineering curricula \cite{buchholz2023}. 
The computer science community has grown in tandem with enrollment, and in the European Union, the information and communication technology sector accounts for 20\% to 50\% of total research and development personnel \cite{eurostat2024}. 
Second, computer science publications represent a substantial share of global research output--roughly one-tenth \cite{ncses2021}--and increased by 21\% from 2015 to 2019, with subfields such as artificial intelligence and robotics experiencing especially rapid growth \cite{unesco2021}. 
Third, public investment in computer science research and education has reached unprecedented levels worldwide, as governments prioritize digital innovation and expand funding for computing and artificial intelligence \cite{unesco2021}. 
For instance, the National Science Foundation in the United States allocated approximately one billion dollars to its Computer and Information Science and Engineering Directorate for fiscal year 2024 \cite{nsf2024}. 
These indicators highlight the discipline's exceptional scale, growth, and influence, making it imperative to examine whether its expansion is accompanied by equitable representation.

Gender imbalance affects many aspects of computer science, including authorship, productivity, recognition, research careers, and academic positions. 
Estimates suggest that women account for only 15--30\% of computer science authors worldwide \cite{holman2018, frachtenberg2022}, with progress increasing at a slow pace \cite{cohoon2011}.
Moreover, the extent of imbalance varies across subfields \cite{frachtenberg2022, peltonen2022}, and female authors face additional disadvantages--such as shorter career lengths--resulting in fewer publications and citations relative to their male counterparts \cite{jadidi2018, huang2020, lietz2024}. 
This imbalance grows more pronounced at advanced career stages, as seen in the United States, where women comprise only 15\% of tenure-track faculty in computer science \cite{clauset2015, way2016}. 
The underrepresentation of women in senior positions may stem from differential retention and advancement rates, with many leaving academia before attaining these ranks \cite{spoon2023, lietz2024}. 
Given the discipline's rapid expansion and global influence, addressing gender inequity through sustained, data-driven investigation is essential and urgent to ensure that the future of computer science is equitable, inclusive, and reflective of its full talent pool.

Research practices--such as authorship, publication venues, and citation behavior--and their links to gender imbalance vary significantly across disciplines.
Recent studies of gendered citation imbalance (i.e., citation imbalance regarding authors' gender) have focused on journal papers within single disciplines, including astronomy \cite{caplar2017}, communications \cite{xwang2021}, dentistry \cite{moreno2024}, economics \cite{dion2018}, environmental studies \cite{oleary2024}, international relations \cite{maliniak2013}, medicine \cite{chatterjee2021}, neuroscience \cite{dworkin2020}, physics \cite{teich2022}, and plant science \cite{pandey2023}. 
However, the extent of such imbalances in computer science remains largely unknown. 
Findings from other disciplines cannot be generalized to computer science, given its unique publishing culture. 
Unlike in most disciplines where journals dominate, computer scientists often present their work at conferences, which involve short peer-review cycles on fixed schedules to expedite dissemination \cite{vardi2009, fortnow2009}. 
Furthermore, conferences and journals are ranked based on citation metrics, with top-ranked venues considered especially prestigious \cite{freyne2010, tsai2014, vrettas2015, li2018}. 
The rapid growth of top-tier conferences has stretched organizing committees beyond capacity, fueling concerns about the thoroughness of peer reviews \cite{vardi2009, fortnow2009, soneji2022}. 
This starkly different publication culture highlights the need to examine how conference-centric practices influence gendered citation patterns.

In this study, we systematically analyze a unique citation network of computer science papers written by authors with assigned binary genders.
We characterize gendered citation patterns as the difference between observed citation behavior and that expected under reference models that account for key structural properties of the network.
This work significantly extends our previous study, which focused exclusively on citations between conference papers in computer science \cite{nakajima2024}, by incorporating journal publications into our dataset and addressing the following new research questions:
(i) What are the patterns of gender imbalance in citations of both conference and journal papers within computer science?
(ii) How does gender imbalance in citations differ between conferences and journals?
(iii) What characteristics of citing papers are associated with gendered citation behavior?

\section{Methods}

\subsection{Construction of a new dataset}

We constructed a unique dataset of 394,432 papers published in computer science conferences and journals, along with 752,742 citations.
By integrating publication data from two open databases, we enriched each paper's metadata with bibliographic details, authors’ binary genders, authorship information, citations, publication venues, and research topics and subfields.
We also identified venue ranks for conferences and journals using external ranking databases.
Because investigating gendered citation behavior requires accounting not only for the gender of authors but also for various author- and paper-level characteristics that may influence citation patterns, this enriched dataset enables a more comprehensive analysis of gendered citation imbalance in computer science than in our previous work \cite{nakajima2024}.
We describe the dataset construction process below.

\subsubsection{DBLP}

The DBLP Computer Science Bibliography (originally an acronym for ``DataBase systems and Logic Programming'', subsequently referred to as DBLP) indexes publications in computer science, encompassing both conferences and journals \cite{dblp, ley2002}. 
DBLP has served as a primary bibliographic resource for computer science and has been widely used for bibliometric analyses on the research landscape of the discipline (e.g., Refs.~\cite{petricek2005, reitz2010, song2014, cavacini2015, kim2019, rosenfeld2023, nakajima2024}).
We used its July 1, 2024 snapshot, which included 7,034,299 papers published in 13,007 conferences and 2,031 journals between 1950 and 2024.
Each paper provides its title, publication year, authors' IDs, authorship order, and the name of the associated conference or journal. 
DBLP employs a proprietary algorithm to assign unique IDs to authors, identifying 3,609,641 authors. 
Previous evaluation showed that DBLP's author name disambiguation performance is highly competitive, possibly due to its hybrid approach combining algorithmic disambiguation and manual error correction \cite{kim2018}.
Each ID consists of the author's full name, optionally followed by a four-digit number if multiple authors share the same name \cite{dblp_author_id}. 
We applied this rule to recover each author's full name from their assigned ID. 
While DBLP offers precise and well-disambiguated venue information, it lacks the broader bibliographic coverage and topic classifications.

\subsubsection{OpenAlex}
OpenAlex provides hundreds of millions of publication records with extensive metadata across multiple disciplines~\cite{priem2022}. 
We used its September 27, 2024 snapshot, which contained 155,449,238 papers published between 1950 and 2024. 
For each paper $v$, we extracted its title, publication date, primary research topic and subfield, authors' names and affiliations, authorship order, and the list of papers cited by $v$.
While every paper lists at least one author, some lack affiliation information for certain authors.
Each paper is classified under one of 4,516 research topics, which are more granular than the 252 subfields to which they belong \cite{openalex_topic_classification}. 
Although OpenAlex provides comprehensive cross-disciplinary coverage of publication and citation data \cite{lorena2024}, we found that many conference and journal names for computer science publications were either missing or not properly disambiguated.

\subsubsection{Matching papers between OpenAlex and DBLP} \label{section:2.1.3}

To reconcile the limitations of both databases, we matched each DBLP paper with its corresponding paper in OpenAlex using three criteria: (i) identical publication year, (ii) matching order of authors' last names, and (iii) a title discrepancy of no more than 25\%, measured by the Levenshtein distance normalized by the longer title.
These criteria have been previously applied to align multiple publication datasets \cite{huang2020}.
The relatively generous 25\% threshold for Levenshtein distance accounts for common discrepancies between databases, such as different encodings of mathematical symbols and the inclusion or exclusion of subtitles.
We treated the last space-separated word in an author's full name as their last name \cite{nakajima2023}.
Using this approach, we identified 3,720,575 DBLP papers with unique matches in OpenAlex.
For each matched paper $v$, we supplemented the DBLP record with OpenAlex metadata, including the title, publication date, primary research topic and subfield, authors' names and affiliations, authorship order, and the list of papers cited by $v$.
Hereafter, we refer to these 3,720,575 enriched papers as $\mathcal{D}$.
There are 23,274,023 citations among papers in $\mathcal{D}$.

\subsubsection{Conference and journal ranks}

We use the following ranking systems to capture the structure of publication prestige in computer science. 
While some resources use a curated list of top-tier venues (e.g., CSRankings \cite{csrankings}), our analysis requires a multi-tiered classification to examine citation imbalances across a broad spectrum of publication venues.

We use the 2021 CORE conference ranking \cite{core_ranking_2021} to determine the ranking of computer science conferences. 
This system evaluates each conference based on its reputation and a range of metrics (e.g., citation statistics of papers and authors, and acceptance rates)\footnote{\url{https://drive.google.com/file/d/1DQixeK53tlq_jh6IspIHroiwu1pmM6-y/edit} (accessed April 2024)}. 
We focus on the 768 conference names assigned to four ranks--$\text{A}^*$, A, B, and C--in which $\text{A}^*$ is the highest and C is the lowest.

Similarly, we use the 2021 SCImago Journal Rank (SJR) \cite{scimago_journal_rank_2021} to rank computer science journals. SCImago draws on Scopus data covering over 34,100 journal titles from more than 5,000 publishers\footnote{\url{https://www.scimagojr.com/aboutus.php} (accessed November 2024)}. 
We restricted our selection to journals indexed under the `Computer Science' subject area for the year 2021. 
This filtering process yielded 1,734 journals assigned to quartiles Q1, Q2, Q3, or Q4, with Q1 being the highest and Q4 the lowest.
While SCImago database indexes journals across disciplines, previous analysis has shown that SJR correlates strongly with the Journal Impact Factor specifically for computer science journals \cite{sicilia2011}.

Because DBLP frequently abbreviates conference and journal names, many entries do not directly match those in the CORE and SCImago data. 
Therefore, for each venue appearing in $\mathcal{D}$, the first and second authors (KN and YS) manually assigned a venue rank ($\text{A}^*$, A, B, C, Q1, Q2, Q3, or Q4) or labeled it ``N/A'' (not applicable). 
This process yielded 60 $\text{A}^*$-ranked, 130 A-ranked, 286 B-ranked, and 233 C-ranked conferences, as well as 341 Q1-ranked, 301 Q2-ranked, 209 Q3-ranked, and 128 Q4-ranked journals.

\subsubsection{Prominence and coauthors of authors} \label{section:2.1.5}

There are 2,430,109 unique author IDs in the papers comprising $\mathcal{D}$. 
For each author $z$, we identify all papers listing $z$ as an author (hereafter, $z$'s papers) and compute two characteristics: (i) prominence, defined as the total number of citations received by $z$'s papers, and (ii) coauthors, the set of authors who have coauthored at least one paper with $z$.
We calculate these metrics using the entire dataset and treat them as static attributes representing the author's prominence and co-authorship networks, which is consistent with our use of static rankings of conferences and journals.

\subsubsection{Country of affiliation and gender of authors}

We assign a country of affiliation to each author based on the affiliations listed in their publications \cite{huang2020, nakajima2023}. 
Specifically, we count the frequency of each country in an author \(z\)'s affiliations and assign the most frequent country if it is unique. 
If no country is uniquely most frequent, we exclude the author from the dataset.
We assigned a country of affiliation to 2,059,647 of the 2,430,109 authors in our dataset.

We then assign binary gender to each author using the method developed in Ref.~\cite{nakajima2023}, which incorporates the author's country of affiliation, first publication year, and first name. 
This method addresses known challenges in name-based gender inference, particularly for East Asian names \cite{nakajima2023}, but is limited to binary classification (see the Discussion Section for further limitations).

To extract first-name candidates from an author's full name, we follow naming conventions based on their country of affiliation. For authors whose country of affiliation is China, Japan, or South Korea and whose name consists of $k$ space-separated words, denoted by $w_1, \ldots, w_k$, we use $w_1$, $w_1 w_2$, $\ldots$, and $w_1 \cdots w_{k-1}$ as $k-1$ candidates of their first name. 
For example, the first-name candidates for the name ``Gil Dong Hong'' are ``Gil'' and ``Gil Dong''. 
For the remaining authors, we assume that the first space-separated word in their name is their sole first-name candidate.

We provide the first-name candidates of an author and their country of affiliation to the Gender API\footnote{\url{https://gender-api.com/en/} (accessed April 2024)}, which was used in previous studies on gender imbalance \cite{dworkin2020, teich2022}. 
The API returns one of three labels (``female,'' ``male,'' or ``unknown'') along with an accuracy score and sample size for each inputted first-name candidate.

We apply thresholds to both the accuracy score and sample size to exclude low-confidence results. 
If multiple first-name candidates exist for an author, we compare the highest female-classified accuracy with the highest male-classified one. 
We assign ``female'' or ``male'' according to which score is higher; if they are equal, no gender is assigned. 
Thresholds were set based on country of affiliation and first publication year to ensure at least 90\% classification accuracy on validation sets with ground-truth gender labels \cite{nakajima2023}. 
Setting different thresholds based on country of affiliation is intended to address potential country imbalances in the gender assignment process, as the precision of name-based gender inference can depend on a name's country of origin \cite{santamaria2018}.
We use the same threshold values as in Ref.~\cite{nakajima2023} (see Supplementary Section S3 in Ref.~\cite{nakajima2023}).

In total, we assigned both country of affiliation and gender to 1,257,016 of the 2,430,109 authors in our dataset.

\subsubsection{Country of affiliation and gender category of papers}

We assign each paper's country of affiliation and gender category following Ref.~\cite{nakajima2023}. 
Specifically, we focus on 1,399,106 papers in $\mathcal{D}$ that satisfy one of two criteria: (i) a single author with an assigned country and gender, or (ii) the first and last authors sharing the same country of affiliation, each with an assigned gender. 
We then assign each paper the country of affiliation of its single author or its first and last authors. 
We also classify each paper into one of four gender categories (``MM,'' ``MW,'' ``WM,'' ``WW'') based on the genders of the first and last authors \cite{dworkin2020, teich2022}.
Specifically, the first letter (M or W) indicates whether the first author is a man or a woman, while the second letter denotes the gender of the last author. 
We categorized sole-authored papers by men or women as MM or WW, respectively. 
We assume that the first and last authors play primary roles in the research and writing of computer science papers, which aligns with common practices in many subfields, although alphabetical ordering is used in some subfields \cite{solomon2009}. 

\subsubsection{Citations}

We focus on the 672,082 papers in $\mathcal{D}$, each assigned a country of affiliation, a gender category (MM, MW, WM, or WW), and a venue rank ($\text{A}^*$, A, B, C, Q1, Q2, Q3, or Q4), and the citations occurring exclusively among these papers. 
Given an observed citation from paper $u$ to paper $v$, we say $u$ ``made'' the citation and $v$ ``received'' it. 
We remove any citation if: (i) $v$'s publication date is ten years older than $u$'s \cite{dworkin2020, nakajima2023}, (ii) $u$ and $v$ share at least one author (self-citation) \cite{nakajima2023}, (iii) the coauthors of $u$'s first or last author overlap with $v$'s authors, or (iv) $u$'s publication year is before 1990 (due to the small number of WW papers before 1980). 
Regarding criterion (i), we followed Refs.~\cite{dworkin2020, nakajima2023}; this ten-year citation window reflects contemporary citation behavior and prevents our analysis from aggregating across disparate eras with substantially different author demographics.
We then discard papers that neither make nor receive any citations, leaving a final set of 394,432 papers and 752,742 citations.
Table~\ref{table:1} summarizes the statistics of our data, comparing the initial matched dataset ($\mathcal{D}$) with the final dataset. 
Note that authors' gender and venue ranks are not assigned in the initial matched dataset $\mathcal{D}$.
We provide additional information on our dataset in Supplementary Section~S1.

\begin{table*}[p]
\caption{Basic statistics of the initial matched dataset ($\mathcal{D}$) and the final dataset.}
 \begin{center}
   \label{table:1}
\begin{tabular}{|l|c|} \hline
\multicolumn{2}{|l|}{\textbf{Initial Matched Dataset}} \\ 
\multicolumn{1}{|c|}{Meaning} & Count \\ \hline
Papers & 3,720,575 \\
Citations & 23,274,023 \\
Authors & 2,430,109 \\
Conferences & 10,735 \\
Journals & 2,004 \\
Countries of affiliation & 214 \\
Research subfields & 243 \\
Research topics & 4,351 \\ \hline \hline
\multicolumn{2}{|l|}{\textbf{Final Dataset}} \\
\multicolumn{1}{|c|}{Meaning} & Count \\ \hline
Papers & 394,432 \\
Citations & 752,742 \\
Gender-assigned authors & 285,406 \\
Conferences & 631 \\
Journals & 955 \\
Countries of affiliation & 140 \\
Research subfields & 238 \\
Research topics & 2,900 \\ \hline
MM papers & 298,063 (75.6\%) \\
MW papers & 30,970 (7.8\%) \\
WM papers & 43,727 (11.1\%) \\
WW papers & 21,672 (5.5\%) \\ \hline
Female authors & 54,229 (19.0\%) \\
Male authors & 231,177 (81.0\%) \\ \hline
$\text{A}^*$-ranked conferences & 60 \\
A-ranked conferences & 124 \\
B-ranked conferences & 254 \\
C-ranked conferences & 193 \\ \hline
Q1-ranked journals & 338 \\
Q2-ranked journals & 294 \\
Q3-ranked journals & 203 \\
Q4-ranked journals & 120 \\ \hline
 \end{tabular}
 \end{center}
\end{table*}

\subsection{Reference models for citation networks}

\begin{table*}[t]
\caption{Properties to be preserved in each reference model. }
 \begin{center}
   \label{table:2}
\begin{tabular}{|l|l|} \hline
   \multicolumn{1}{|c|}{Model} & \multicolumn{1}{c|}{Properties to be preserved} \\ \hline
   Random-draws & $\bullet$ Number of citations made by each paper \\ \hline
   \multirow{2}{*}{Homophilic-draws} & $\bullet$ Number of citations made by each paper \\
    & $\bullet$ Homophily in citation patterns \\ \hline
   \multirow{3}{*}{Preferential-draws} & $\bullet$ Number of citations made by each paper \\
   & $\bullet$ Homophily in citation patterns \\
   & $\bullet$ Heterogeneity in the number of citations received per paper\\
   \hline
 \end{tabular}
 \end{center}
\end{table*}

Reference models randomize the citations made by each paper while preserving certain structural properties of the original network. 
Dworkin et al.~introduced the random-draws model as a reference model \cite{dworkin2020}, but it preserves only the number of citations made by each paper, which limits its ability to capture how other network properties contribute to gender imbalance in citations. 
To address this gap, we propose a family of reference models that preserve the number of citations made by each paper and up to two additional properties \cite{nakajima2023}: homophily in citations \cite{ciotti2016, zeng2017} and heterogeneity in the number of citations received per paper \cite{eom2011, zeng2017}. 
Table~\ref{table:2} outlines the properties each model preserves.
In the following, we denote by \(V = \{v_1, \ldots, v_N\}\) the set of all the papers, \(E\) the set of citations in the network, and \(M = |E|\) the number of citations, where \(N\) is the number of papers.

\subsubsection{Random-draws model}

We first describe the random-draws model \cite{dworkin2020}. 
Let \(\overline{V}_{\text{RD}}(i)\) denote the set of papers that paper \(v_i \in V\) could potentially cite under this model. 
Following our citation criteria, \(\overline{V}_{\text{RD}}(i)\) comprises papers that (i) are at most ten years older than \(v_i\), (ii) do not share any author with \(v_i\), and (iii) do not share any author with the coauthors of \(v_i\)'s first or last author. 
For each citation \((v_i, v_j) \in E\), we randomly draw \(v_{j'} \in \overline{V}_{\text{RD}}(i)\) with uniform probability and replace \((v_i, v_j)\) by \((v_i, v_{j'})\) with replacement. 
The random-draws model preserves the number of citations made by each paper in the original network.

\subsubsection{Homophilic-draws model}

We extend the random-draws model to preserve homophily in citations, referring to this model as the homophilic-draws model. 
Let \(S\) represent a set of paper characteristics (beyond gender category) potentially relevant for homophily in citations. 
In this study, we focus on (i) country of affiliation, (ii) primary research topic, and (iii) venue rank. 
Note that \(S\) is not limited to these attributes and may vary across disciplines.

For each citation \((v_i, v_j) \in E\), let \(\overline{V}_{\text{HD}}(i, j, S)\) be the subset of \(\overline{V}_{\text{RD}}(i)\) (from the random-draws model) consisting of papers that share all attributes in \(S\) with \(v_j\).
Note that $\overline{V}_{\text{HD}}(i, j, S)$ includes $v_{j}$. 
In our analyses, \(\overline{V}_{\text{HD}}(i, j, S)\) contains papers matching \(v_j\) in country of affiliation, research topic, and venue rank. 
We replace each citation \((v_i, v_j)\) by \((v_i, v_{j'})\) with replacement, where \(v_{j'}\) is uniformly drawn from \(\overline{V}_{\text{HD}}(i, j, S)\). 
The homophilic-draws model preserves the number of citations made by each paper and the number of citations between pairs of countries of affiliation, topics, and venue ranks.

\subsubsection{Preferential-draws model}

We further extend the homophilic-draws model to preserve, approximately, the heterogeneity in citations received per paper. 
We refer to this as the preferential-draws model, which randomizes the citations made by each paper while respecting the characteristics in \(S\) and preferential attachment dynamics.

First, we sort the papers by publication date in ascending order, denoted \(\{v_{x_1}, \ldots, v_{x_N}\}\), where \(v_{x_1}\) is the oldest and \(v_{x_N}\) is the newest. Let \(c_{i, l, \text{PD}}\) be the number of citations received by paper \(v_i\) from papers \(v_{x_1}, \ldots, v_{x_l}\) in the model. 
We initialize \(c_{i, l, \text{PD}} = 0\) for all \(i\) and \(l\). 
For each citation \((v_{x_l}, v_j) \in E\), we define
\[
\overline{V}_{\text{PD}}(x_l, j, S) 
= \bigl\{\,v_{j'} \in \overline{V}_{\text{HD}}(x_l, j, S) \,\big|\,
\lfloor \ln(c_{j', l-1, \text{PD}}+1) \rfloor 
= \lfloor \ln(c_{j, l-1, \text{PD}}+1) \rfloor \bigr\},
\]
where \(\lfloor x \rfloor\) is the floor function. 
Thus, \(\overline{V}_{\text{PD}}(x_l, j, S)\) contains papers in \(\overline{V}_{\text{HD}}(x_l, j, S)\) that have received roughly the same number of citations as \(v_j\) from \(\{v_{x_1}, \ldots, v_{x_{l-1}}\}\).
We use the natural logarithm to group papers into bins corresponding to a logarithmic scale of the number of citations received. 
This log-binning accounts for a heavy-tailed distribution of the number of citations received by a paper and prevents the candidate pool (i.e., $\overline{V}_{\text{PD}}(x_l, j, S)$) from becoming too sparse while approximately preserving the number of citations received by the paper being replaced.

We then sequentially randomize the citations made by \(v_{x_l}\) from \(l=2\) to \(N\). 
For each citation \((v_{x_l}, v_j) \in E\), we randomly select \(v_{j'}\) from \(\overline{V}_{\text{PD}}(x_l, j, S)\) with uniform probability, replacing \((v_{x_l}, v_j)\) by \((v_{x_l}, v_{j'})\) with replacement. 
Note that \(v_{x_1}\) makes no citations, as it is the oldest paper.
This sequential process allows the citation count (i.e., $c_{j, l-1, \text{PD}}$) to update dynamically and approximates the preferential attachment dynamics, where papers that accumulate citations early in the process become more likely to be cited by subsequent papers.

Like the homophilic-draws model, this preferential-draws model preserves the number of citations made by each paper and homophily in citations with respect to the country of affiliation, research topic, and venue rank. 
In addition, it approximately preserves the original distribution of the number of citations received per paper. 
See Supplementary Section S2 for a pseudocode of the preferential-draws model. 
See Appendix \ref{appendix:a} for a comparison of structural properties among the three reference models.

\subsection{Characterizing gender imbalance in citations}

A standard practice in network analysis is to compare the original network with randomized instances generated by reference models in terms of structure and dynamics \cite{newman2001, milo2002, orsini2015, fosdick2018}.
We characterize gender imbalance in citations by the difference in citation patterns regarding authors' gender between the original network and reference models.
Consider two subsets of papers, denoted by $V_{\text{from}}$ and $V_{\text{to}}$. 
We measure the extent to which the papers in $V_{\text{from}}$ over- or under-cite the papers in $V_{\text{to}}$ and in a given gender category $g \in \{\text{MM}, \text{MW}, \text{WM}, \text{WW}\}$ \cite{dworkin2020, teich2022}.
We first count the citations made by the papers in $V_\text{from}$ to those in $V_{\text{to}}$ and in $g$ in the original network, which we denote by $n_{g, \text{obs}}$. 
Then, we independently generate 100 randomized networks using the reference model.
We denote by $\mu_{g, \text{rand}}$ and $\sigma_{g, \text{rand}}$ the mean and standard deviation of the number of citations made by the papers in $V_\text{from}$ to those in $V_{\text{to}}$ and in $g$ across the 100 randomized instances.
We define the over/under-citation made by the papers in $V_{\text{from}}$ to those in $V_{\text{to}}$ as $(n_{g, \text{obs}} - \mu_{g, \text{rand}}) / \mu_{g, \text{rand}}$ \cite{dworkin2020, teich2022}.

To test the statistical significance of the over/under-citation, we calculate the Z score defined by $(n_{g, \text{obs}} - \mu_{g, \text{rand}}) / \sigma_{g, \text{rand}}$. 
We say that the papers in $V_{\text{from}}$ significantly over-cite those in $V_{\text{to}}$ if the Z score is larger than 3.09 (i.e., $p$ value is lower than 0.001). 
Similarly, we say that the papers in $V_{\text{from}}$ significantly under-cite those in $V_{\text{to}}$ if the Z score is lower than $-3.09$ (i.e., $p$ value is lower than 0.001).

\subsection{Matched-pair analysis of gender imbalance in citations}

\begin{table*}[t]
\caption{Number of matched paper pairs used in the analysis.  
Each row shows a pair of paper subsets \(\mathcal{M}\) and \(\mathcal{M}'\), and the number of matched pairs generated between them. }
 \begin{center}
   \label{table:3}
\begin{tabular}{|l|l|c|} \hline
   \multicolumn{1}{|c}{\multirow{2}{*}{$\mathcal{M}$}} & \multicolumn{1}{|c}{\multirow{2}{*}{$\mathcal{M}'$}} & \multicolumn{1}{|c|}{Matched} \\ 
   & & pairs \\ \hline
   WW papers & MM papers & 10,985 \\ \hline
   MM (conference) & MM (journal) & 49,860 \\ 
   MM (with prominent authors) & MM (without prominent authors) & 29,290 \\ 
   MM (top half of $\text{MA}_{\text{or}}$) & MM (bottom half of $\text{MA}_{\text{or}}$) & 50,124 \\ \hline
   WW (conference) & WW (journal) & 1,215 \\ 
   WW (with prominent authors) & WW (without prominent authors) & 741 \\ 
   WW (top half of $\text{MA}_{\text{or}}$) & WW (bottom half of $\text{MA}_{\text{or}}$) & 1,166 \\ 
   \hline
 \end{tabular}
 \end{center}
\end{table*}

To examine which types of papers contribute to gender imbalance in citations, we compare over/under-citation patterns while controlling for key paper-level characteristics \cite{dworkin2020, teich2022}.
We adopt a matched-pair analysis, as used in previous studies on gender imbalance \cite{huang2020, li2022}, to identify which characteristics of papers that made citations are associated with stronger gender imbalance in their citation behavior.

Let \(\mathcal{M}\) and \(\mathcal{M}'\) be two disjoint subsets of papers.
We generate a set of matched pairs as follows. 
For each paper \(u \in \mathcal{M}\), we randomly select (without replacement) a paper \(v \in \mathcal{M}'\) that matches $u$ on the following characteristics: (i) publication year, (ii) country of affiliation, (iii) primary research subfield, and (iv) number of citations made by paper.
We repeat this process 100 times to generate 100 independent sets, denoted by \(\{(\mathcal{M}, \mathcal{M}'_i)\}_{i=1}^{100}\), where \(\mathcal{M}'_{i} \subseteq \mathcal{M}'\) for \(i=1,\ldots,100\). 
 We compute \(\Delta_{\mathcal{M}, g}\), the over/under-citation made by papers in \(\mathcal{M}\) to papers in a gender category \(g \in \{\text{MM}, \text{MW}, \text{WM}, \text{WW}\}\). 
Similarly, for each \(i=1,\ldots,100\), we compute \(\Delta_{\mathcal{M}'_i, g}\), the over/under-citation made by papers in \(\mathcal{M}'_{i}\) to papers in \(g\). 
Let \(\bar{\Delta}_{\mathcal{M}', g}\) and \(s_{\mathcal{M}', g}\) denote the mean and standard deviation, respectively, of these 100 values.
The test statistic is defined as
\begin{align}
t = \frac{\bar{\Delta}_{\mathcal{M}', g} - \Delta_{\mathcal{M}, g}}{s_{\mathcal{M}', g} / \sqrt{100}}
\label{eq:4}
\end{align}
We test the null hypothesis that \(\Delta_{\mathcal{M}, g} = \bar{\Delta}_{\mathcal{M}', g}\), against the alternative that they differ. 
If the resulting \(p\)-value is below 0.001, we reject the null hypothesis in favor of the alternative.

We first test whether the gender of citing authors is associated with citation imbalance using matched MM--WW paper pairs. 
Next, we examine three additional characteristics of citing papers: (i) venue type (conference versus journal), (ii) involvement of ``prominent'' authors, and (iii) local coauthorship network. 
For the characteristic (ii), we define prominent authors as the top 1\% of each gender by author prominence. 
For the characteristic (iii), we use the man-author over-representation (\(\text{MA}_{\text{or}}\)), defined as the proportion of male coauthors of the first or last author of a paper \(u\), minus the overall proportion of male authors in the dataset \cite{dworkin2020}.
See Section \ref{section:2.1.5} for the definitions of authors' prominence and coauthors.
For each characteristic, we split the MM papers into two disjoint sets--specifically, MM conference versus~journal papers; MM papers with a prominent first or last author versus those with neither; and MM papers in the top versus bottom half of \(\text{MA}_{\text{or}}\). 
We then generate matched pairs between the two sets. 
For WW papers, we apply the same procedure as for MM papers.
Table~\ref{table:3} reports the number of matched paper pairs used in our analysis, including MM--WW comparisons and within-group comparisons based on venue type, involvement of prominent authors, and local coauthorship network. 

\section{Results} \label{section:2}

\subsection{Quantifying gendered citation imbalance in computer science}

\begin{figure*}[t]
\centering
\includegraphics[width=1.0\textwidth]{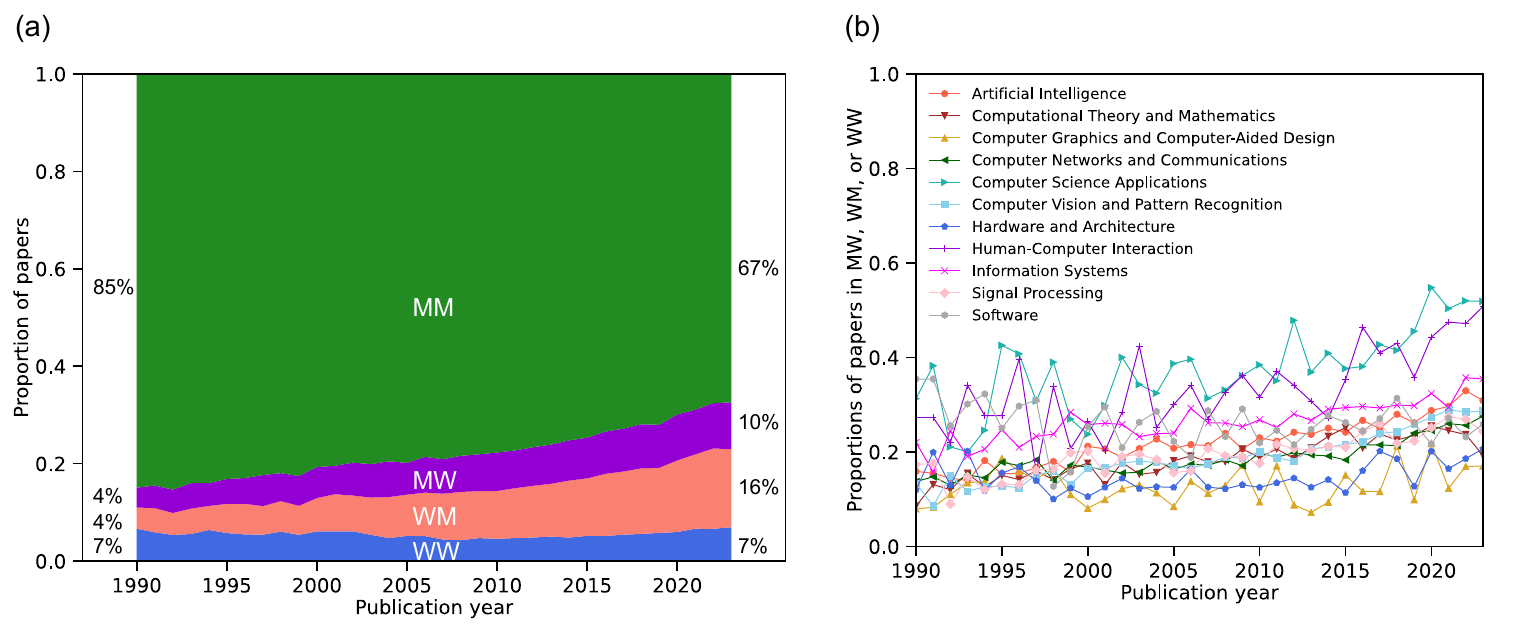}
\caption{
Time-varying demographics of published papers by gender category in computer science. (a) Proportions of papers by gender category over time. (b) Proportions of papers in gender categories MW, WM, or WW across 11 subfields of computer science over time.
}
\label{fig:1}
\end{figure*}

We constructed a citation network comprising 394,432 papers published in computer science conferences and journals, along with 752,742 citations between them. 
We classified each paper into one of four gender categories--MM, MW, WM, or WW--based on the genders of its first and last authors. 
Here, the first letter (M or W) denotes the gender of the first author, and the second letter that of the last author. 
Sole-authored papers by men or women are categorized as MM or WW, respectively.
Our dataset includes 298,063 MM papers (75.6\%), 30,970 MW papers (7.8\%), 43,727 WM papers (11.1\%), and 21,672 WW papers (5.5\%). 
Between 1990 and 2023, the share of papers with women as the first and/or last authors increased from 15\% to 33\% (Fig.~\ref{fig:1}(a)). 
Although the pace of increase varies across subfields, the overall trend is upward (Fig.~\ref{fig:1}(b)).

\begin{figure*}[t]
\centering
\includegraphics[width=0.92\textwidth]{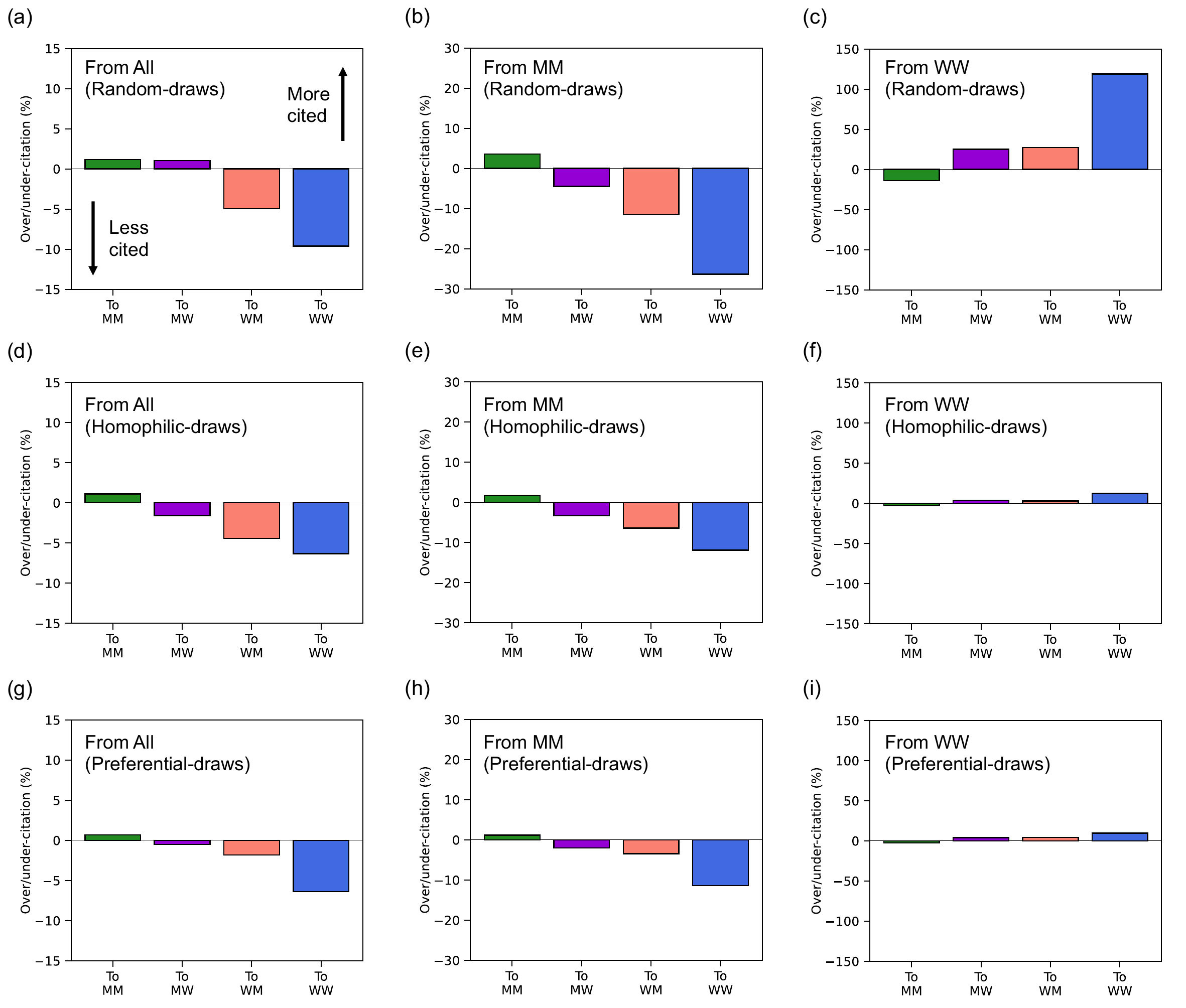}
\caption{
Gender imbalance in citations made by papers in computer science. Panels (a), (d), and (g) show results for all papers; (b), (e), and (h) for MM papers; and (c), (f), and (i) for WW papers. We used the random-draws model in (a)--(c), the homophilic-draws model in (d)--(f), and the preferential-draws model in (g)--(i). See Supplementary Tables S6--S8 for the statistical significance of the over/under-citation under these three reference models.
}
\label{fig:2}
\end{figure*}

\begin{figure*}[p]
\centering
\includegraphics[width=1.0\textwidth]{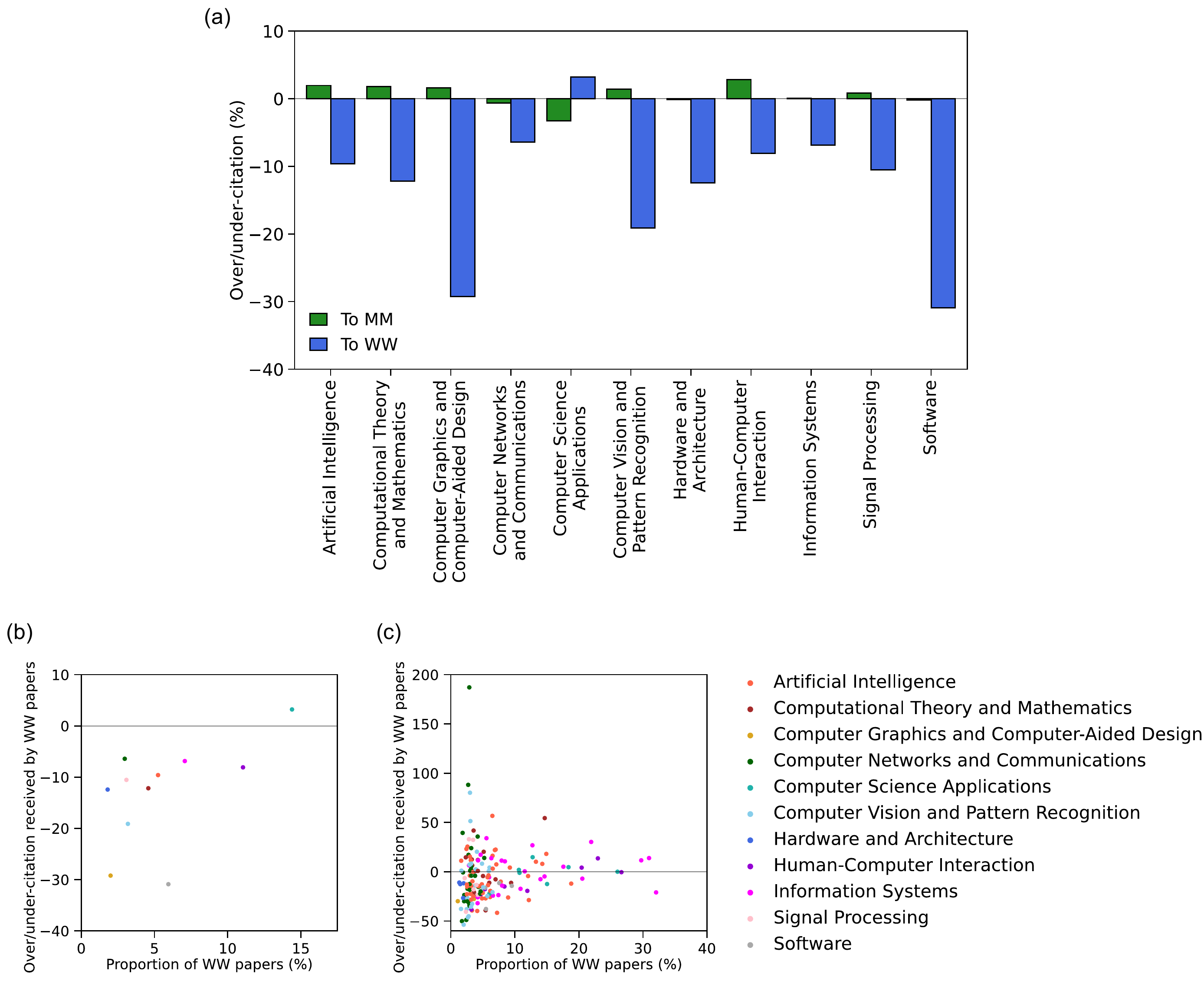}
\caption{
Gendered citation imbalance across subfields and topics in computer science. 
(a) Gender imbalance in citations made by all papers to MM and WW papers in each subfield of computer science. 
See Supplementary Table S9 for the statistical significance of the over/under-citation.
(b) Proportion of WW papers in a subfield versus the over/under-citation made by all papers to these papers.  
(c) Proportion of WW papers in a topic versus the over/under-citation made by all papers to these papers.
}
\label{fig:3}
\end{figure*}

\begin{figure*}[p]
\centering
\includegraphics[width=1.0\textwidth]{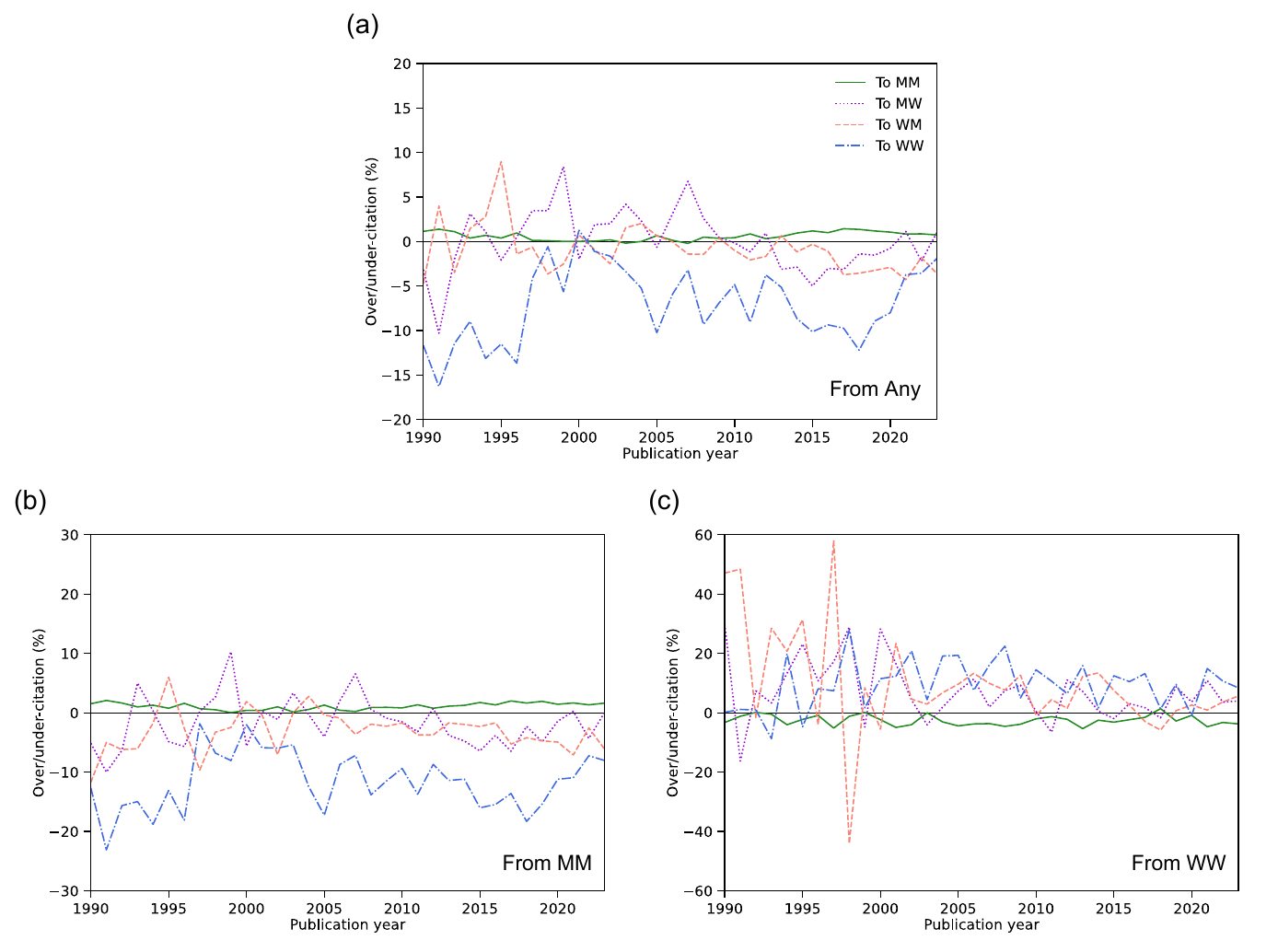}
\caption{
Temporal trends of gender imbalance in citations made by papers in computer science. (a): All papers. (b): MM papers. (c): WW papers. See Supplementary Figure S3 for the statistical significance of the over/under-citation over time}.
\label{fig:4}
\end{figure*}

We now quantify gendered citation imbalance in computer science. 
We first count the number of citations made by a subset of papers to papers in each gender category.
Then, we compare the obtained citation counts with the expected numbers derived from a reference model that randomizes citations while preserving certain properties of the original network.
We define the ``over/under-citation'' as (observed $-$ expected)/expected, which captures how much the observed citation count exceeds (or falls short of) the expected value (see Methods).
This approach adopts a standard practice for characterizing subgraphs or patterns appearing in empirical networks and will reveal aspects of citation imbalance not immediately explained by common network properties \cite{uzzi2013, orsini2015, fosdick2018, kojaku2021}.

We begin with the random-draws model \cite{dworkin2020}, which preserves the number of citations made by each paper, as the reference model.
Under this model, we found that all papers in our dataset significantly over-cite MM papers and under-cite WM and WW papers (see Fig.~\ref{fig:2}(a)).
We further categorize citations made by papers by gender category \cite{dworkin2020, teich2022}; unless we state otherwise, we focus on citations made by MM and WW papers to distinguish how papers authored by men and by women differ in their citation practices.
MM papers significantly over-cite MM papers and under-cite MW, WM, and WW papers (see Fig.~\ref{fig:2}(b)). 
In contrast, WW papers significantly under-cite MM papers and over-cite MW, WM, and WW papers (see Fig.~\ref{fig:2}(c)). 
These results are consistent with previous studies on gender imbalance in citations among journal papers in neuroscience and physics \cite{dworkin2020, teich2022}.

Citation networks exhibit two common properties \cite{zeng2017}: homophily in citations--where papers tend to cite others with similar characteristics--and heterogeneity in the number of citations received per paper--where most papers receive few citations while a small fraction accumulate many. 
Because these properties often drive structural and dynamical behaviors in citation networks \cite{zeng2017}, we hypothesize that they contribute to gendered citation imbalance. 
To test this hypothesis, we extend the random-draws model, which preserves only the number of citations made by each paper but destroys both homophily and citation heterogeneity, to a family of reference models. 
The homophilic-draws model preserves homophily with respect to (i) country of affiliation \cite{gomez2022, nakajima2023}, (ii) research subfield and topic \cite{tekles2022}, and (iii) venue rank \cite{tsai2014, li2018}, while the preferential-draws model additionally preserves heterogeneity in the number of citations received per paper. 
Comparing these models allows us to examine the respective impacts of these two properties on gendered citation imbalance.

Homophily in citations contributes to the over/under-citation made by all papers.
Indeed, the over/under-citation received by WW papers decreases from $-10.1\%$ in the random-draws model (see Fig.~\ref{fig:2}(a)) to $-6.3\%$ in the homophilic-draws model (see Fig.~\ref{fig:2}(d)).
We observe a similar trend for the over/under-citation made by MM papers (see Figs.~\ref{fig:2}(b) and \ref{fig:2}(e)); furthermore, the over/under-citation made by WW papers is largely explained by homophily in citations (see Figs.~\ref{fig:2}(c) and \ref{fig:2}(f)).
This indicates that the gendered citation pattern made by WW papers may largely be attributable to structural factors, such as the concentration of female researchers in certain combinations of topic, country, and venue.
By contrast, the heterogeneity in the number of citations received per paper has little impact. 
In fact, the over/under-citation made by all papers is quantitatively comparable between the homophilic-draws and preferential-draws models (see Figs.~\ref{fig:2}(d) and \ref{fig:2}(g)).
We observe similar trends for the over/under-citation made by MM papers (see Figs.~\ref{fig:2}(e) and \ref{fig:2}(h)) and by WW papers (see Figs.~\ref{fig:2}(f) and \ref{fig:2}(i)).

To summarize, a persistent gendered citation imbalance remains in computer science, even after accounting for both homophily in citations and heterogeneity in the number of citations received per paper. 
Homophily in citations is strongly associated with this imbalance, while the latter has a minor impact. 
Our findings suggest that (i) addressing excessive homophily in citing practices, including over-citations among researchers from the same country of affiliation \cite{baccini2023}, is crucial, and (ii) simply focusing on highly-cited papers is unlikely to eliminate the imbalance.

Unless we state otherwise, we employ the preferential-draws model in the subsequent analysis. 
This model quantifies gender imbalance in citations that is not immediately explained by the number of citations made by each paper, homophily in citations, or heterogeneity in the number of citations received per paper.
While we exclude self-citations here, the gender imbalance is more pronounced when we include (see Supplementary Section S4 for details), which is largely consistent with previous results on gendered self-citation behaviors \cite{king2017, dworkin2020, teich2022}.
Moreover, while we excluded isolated papers (i.e., those that neither make nor receive any citations) from our analyses, a sensitivity analysis confirmed that our main findings on over/under-citation patterns remain qualitatively unchanged (see Supplementary Section S5 for details).

Because the extent of gender imbalance may vary across computer science subfields \cite{frachtenberg2022, peltonen2022}, we analyze the 11 subfields of computer science classified in OpenAlex.
We compute the over/under-citation made by all papers to MM and WW papers in each subfield and compare these values across subfields.
As expected, the degree of gendered citation imbalance varies among them (Fig.~\ref{fig:3}(a)).
These variations may be associated with gender differences in research focus; recent evidence indicates that women are more represented in applied research areas, yet faculty rated researchers in these areas as less likely to publish, receive tenure, or obtain funding compared to theoretical ones \cite{kleinberg2025}. Such differences in research focus could act as a confounding factor for the observed subfield-level citation imbalance.
Figure~\ref{fig:3}(b) plots the proportion of WW papers in each subfield (horizontal axis) against the over/under-citation made by all papers to those papers (vertical axis).
We find a moderate positive correlation between these two measures (Spearman's rank correlation coefficient = $0.455$).

A similar pattern of correlation between the proportion of WW papers and over/under-citation is observed at the topic level.
Specifically, we consider the 172 topics within computer science subfields that each contain at least 10 WW papers and have collectively received at least 10 citations to those papers.
Figure~\ref{fig:3}(c) shows the proportion of WW papers in each topic versus the over/under-citation made by all papers to those papers.
We observe substantial variation in over/under-citation across research topics and a weaker positive correlation (Spearman's rank correlation coefficient = $0.202$).

Although the proportion of WW papers published each year has steadily increased (see Fig.~\ref{fig:1}(a)), the over-citation received by MM papers and under-citation received by WW papers exhibit relatively stable trends over time (see Fig.~\ref{fig:4}(a)). 
When we categorize papers by gender each year, the temporal trend in over/under-citation made by MM papers closely mirrors that made by all papers (see Fig.~\ref{fig:4}(b)). 
WW papers consistently under-cite MM papers and over-cite WW papers, and this pattern has remained similarly stable (see Fig.~\ref{fig:4}(c)). 
These findings align with the previous results for neuroscience and physics disciplines \cite{dworkin2020, teich2022} and suggest that citation practices become fixed over the long term and do not easily adapt to a diversifying author population.

\subsection{Conferences versus journals}

\begin{figure*}[p]
\centering
\includegraphics[width=1\textwidth]{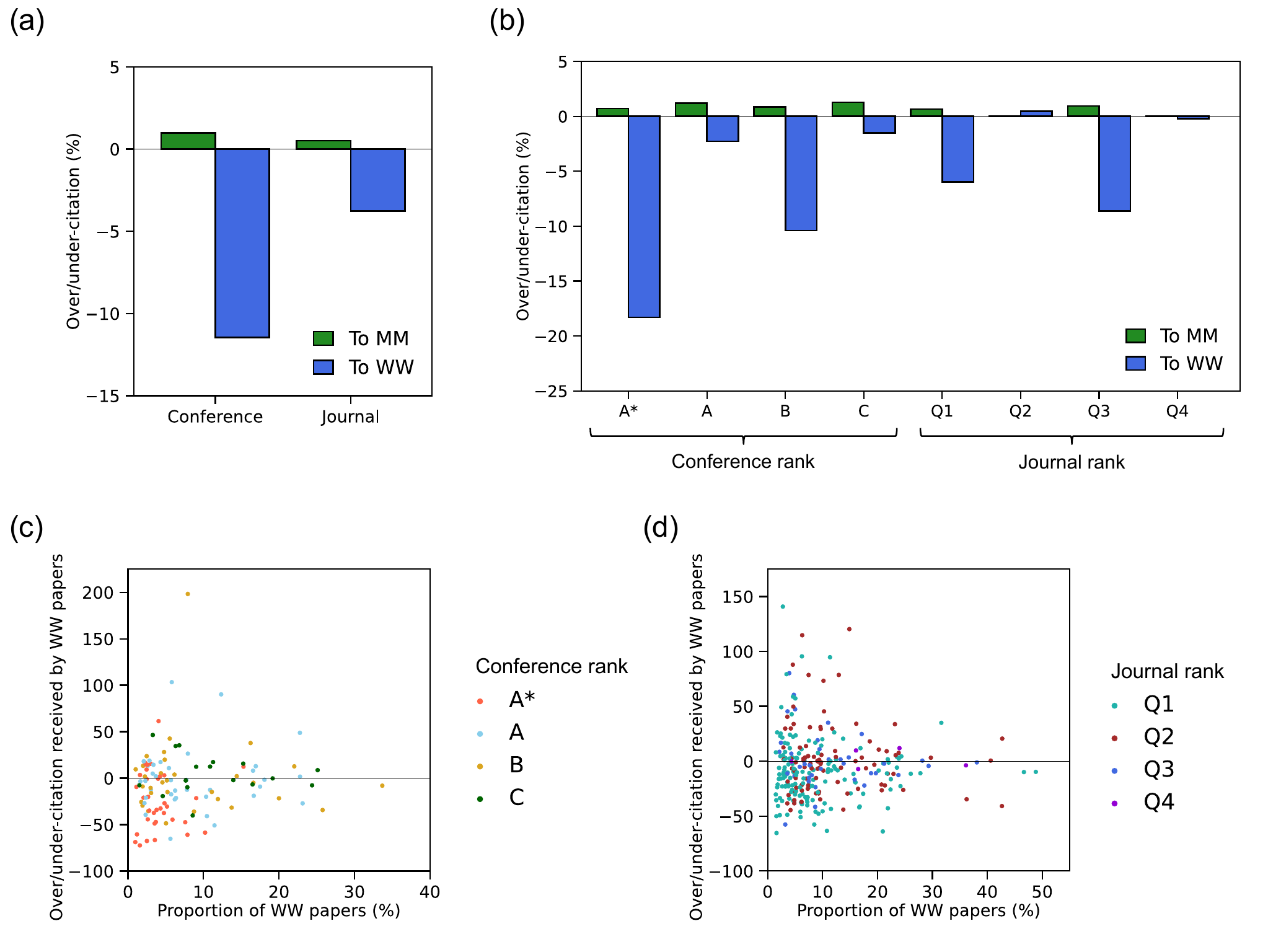}
\caption{
Gendered citation imbalance in different publication venues of computer science. 
(a) Over/under-citation received by MM and WW papers published in conferences and journals. See Supplementary Table S10 for the statistical significance of the over/under-citation.
(b) Over/under-citation received by MM and WW papers in each venue rank. See Supplementary Table S11 for the statistical significance of the over/under-citation. 
(c) Proportion of WW papers in a conference versus over/under-citation received by these papers.
(d) Proportion of WW papers in a journal versus over/under-citation received by these papers.
}
\label{fig:5}
\end{figure*}

In computer science, international conferences often serve as the primary publication venue, filling a role that journals typically occupy in other disciplines \cite{freyne2010, franceschet2010, vrettas2015}. 
Their distinctive publishing culture, characterized by short review cycles and other unique features, may shape authors' citation practices. 
Indeed, a longstanding debate has compared conferences and journals in this discipline \cite{vardi2009, fortnow2009, franceschet2010, vrettas2015, kim2019}.
Here, we examine conferences and journals through the lens of gendered citation imbalance.

We compare the over/under-citation made by all papers to conference versus journal papers in a gender category (MM or WW). 
MM conference papers are over-cited at twice the rate of MM journal papers, while WW conference papers are under-cited at three times the rate of WW journal papers (Fig.~\ref{fig:5}(a)). 
This result suggests that authors citing conference papers may reference a narrower range of researchers, potentially contributing to the observed gendered citation imbalance in computer science.
Furthermore, this disparity suggests that the field's unique publication practices, in which researchers often present their work at international conferences rather than in journals \cite{vardi2009, fortnow2009}, may create a prestige hierarchy among publication venues that intersects with gendered citation imbalance. 
This observation largely aligns with previous studies on the intersection of gender imbalance and prestige regarding research institutions, research focus, and publication venues \cite{way2016, bendels2018, huang2020, kleinberg2025}.

Conferences and journals in computer science are frequently ranked according to citation metrics of the papers they publish \cite{freyne2010}. 
The highest-ranked venues are viewed as especially prestigious \cite{tsai2014, li2018}. 
We explore relationships between the venue rank and the extent of gendered citation imbalance.
One common conference ranking is the CORE ranking system, which designates the top conferences as $\text{A}^*$, followed by A, B, and C. 
For journals, we use the SCImago Journal Rank, which classifies the top journals as Q1, followed by Q2, Q3, and Q4. 
We compute the over/under-citation made by all papers to papers that belong to each venue rank and gender category (MM or WW).

Venue prestige is associated with gendered citation imbalance in computer science, and this association is more pronounced for conferences than for journals (Fig.~\ref{fig:5}(b)). 
MM papers are over-cited in all conference tiers and in Q1 and Q3 journals, whereas WW papers are under-cited in $\text{A}^*$ and B conferences as well as Q1 and Q3 journals. 
WW papers in $\text{A}^*$-ranked conferences are under-cited at three times the rate observed in Q1 journals, whereas the over-citation received by MM papers in $\text{A}^*$-ranked conferences is on par with that in Q1 journals.

Gendered citation imbalance is less pronounced, or even reversed, in middle-tier venues (Q2 journals and A-ranked conferences) compared to the top-tier venues (Q1 journals and $\text{A}^*$-ranked conferences) 
This non-monotonic pattern aligns with previous findings suggesting that prestige bias is often most acute at the top tier of the hierarchy. 
For instance, authors from top-ranked institutions often benefit disproportionately from their affiliation prestige, whereas those from middle-tier or lower-ranked institutions may experience different evaluation dynamics in peer-review processes \cite{blank1991, sun2022, hultgren2024}.
Similarly, our results suggest that the gendered citation imbalance can be amplified in top-tier conferences and journals, potentially due to their greater visibility and competition relative to middle-tier venues.

We examine the variation in over/under-citation received by WW papers across the 120 conferences and 292 journals, each meeting two criteria: (i) publication of at least 10 WW papers and (ii) at least 10 total citations received by those WW papers. 
In both conference and journal, the proportion of WW papers published in the venue shows almost no correlation with the over/under-citation received by these papers; Spearman's rank coefficients of 0.092 for conferences (Fig.~\ref{fig:5}(c)) and 0.026 for journals (Fig.~\ref{fig:5}(d)).

To summarize, gendered citation imbalance is evident in both conferences and journals within computer science, but it is more pronounced for conference papers, especially those published at top-tier conferences. 
Furthermore, the variation in the extent of gendered citation imbalance across conferences and journals is largely unexplained by the underlying gender imbalance in presenting authors.
This disparity between presenting authors' gender ratios and actual citation practices may stem from systemic factors, such as conference-dominant culture \cite{vardi2009, fortnow2009}, the involvement of prominent authors \cite{tomkins2017, sun2022}, and coauthorship networks \cite{menezes2009, kim2019}.

\subsection{Potential drivers of gendered citation imbalance}

\begin{figure*}[p]
\centering
\includegraphics[width=1.0\textwidth]{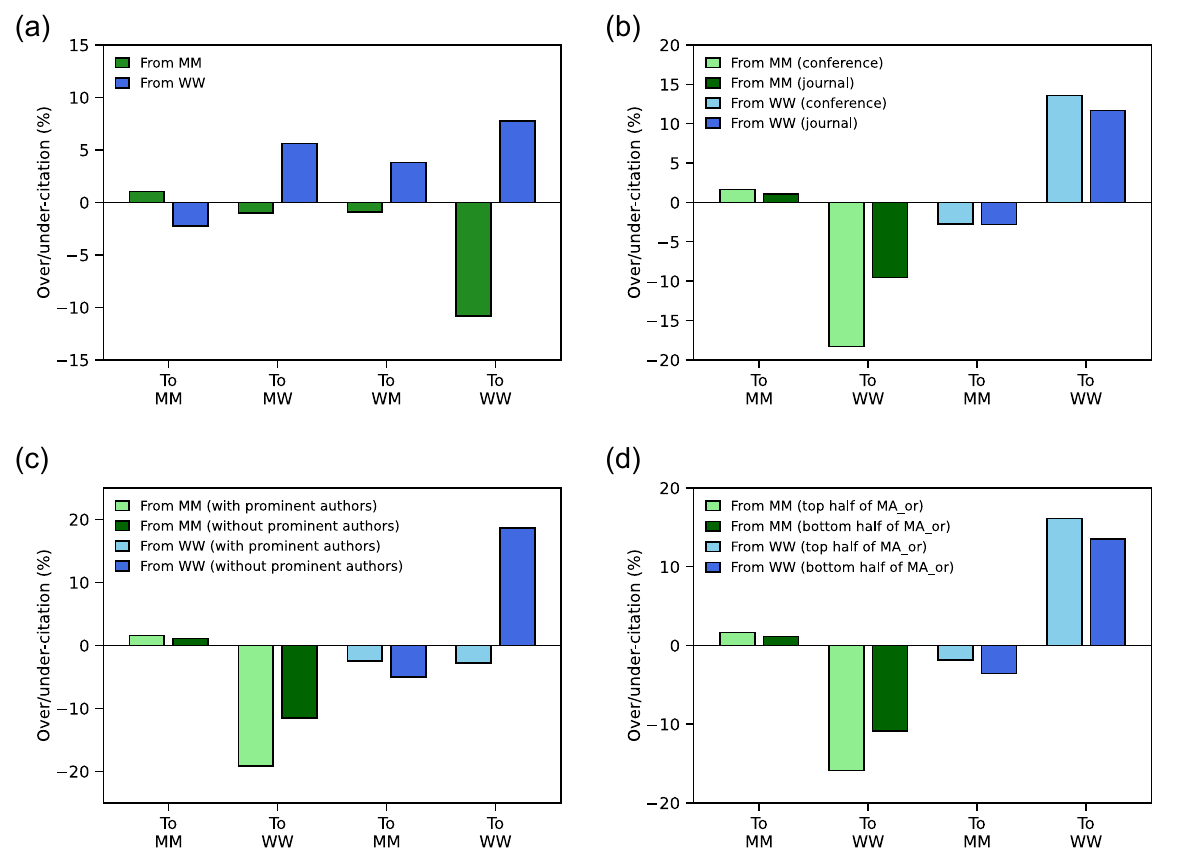}
\caption{
Matched pair analysis on potential drivers of gender imbalance in citations made by papers. (a) Authors' gender. (b) Venue type. (c) Involvement of prominent authors. (d) Local coauthorship network.
See Supplementary Tables S14--S17 for the statistical significance of the over/under-citation shown in panels (a), (b), (c), and (d).
}
\label{fig:6}
\end{figure*}

Based on these findings, we examine potential drivers of gendered citation imbalance in computer science. 
To do so, we compare over/under-citation patterns made by papers while controlling for key paper characteristics \cite{dworkin2020, teich2022}. 
We employ a matched-pair analysis, as in previous studies on gender imbalance in research careers \cite{huang2020, li2022}, to assess whether certain characteristics of papers that made citations are associated with stronger gender imbalance in their citation behavior.

We first hypothesize that an author's gender influences the gender imbalance in citations made by their papers, as suggested by the patterns shown in Fig.~\ref{fig:2}. 
To test this, we conduct a matched-pair analysis that controls for paper characteristics beyond gender category (see Methods). 
We generate independent sets of matched MM and WW paper pairs--each pair matching in publication year, country of affiliation, subfield, and number of citations made--and then compare the over/under-citation made by MM versus WW papers across these matched sets.

The over/under-citation made by MM papers to papers in each gender category differs significantly from that made by WW papers (Fig.~\ref{fig:6}(a)). 
Specifically, MM papers over-cite MM and under-cite MW, WM, and WW, whereas WW papers under-cite MM and over-cite MW, WM, and WW. 
These findings align with Figs.~2(h) and 2(i), and these gendered citation patterns remained after controlling publication year, country of affiliation, subfield, and number of citations made by the paper.

We next examine how specific characteristics of MM and WW papers contribute to gender imbalance in citations made by these papers. 
We focus on three characteristics: (i) venue type (conference or journal), (ii) involvement of ``prominent'' authors (those with highly cited papers), and (iii) local coauthorship network. 
Using a matched-pair analysis again, we test whether each characteristic influences gender imbalance (see Methods). 
First, we split the MM (or WW) papers into two disjoint sets based on a given characteristic; for instance, MM conference papers versus MM journal papers in the case of venue type. 
Second, we generate independent sets of matched MM (or WW) pairs, controlling for publication year, country of affiliation, subfield, and the number of citations made by the paper. 
Finally, we compute the over-/under-citation made by MM papers (or WW papers) across the matched sets to assess the association of that characteristic with gendered citation imbalance.

We identified several associations between these three characteristics and the gender imbalance in citations by MM and WW papers. 
First, the absolute magnitude of over/under-citation made by MM conference papers exceeds that of MM journal papers (Fig.~\ref{fig:6}(b)). 
While the under-citation to MM papers remains comparable between WW conference papers and WW journal papers, the over-citation to WW papers is larger in WW conference papers than in WW journal papers (Fig.~\ref{fig:6}(b)). 
Second, the involvement of prominent authors exhibits qualitatively different effects for MM and WW papers (Fig.~\ref{fig:6}(c)): the absolute magnitude of the over/under-citation made by MM papers is greater when prominent authors are involved, whereas that made by WW papers is smaller. 
This suggests that while the author's prominence amplifies male homophily, it diminishes female homophily. This result may indicate that prominent female authors could adopt the dominant, male-centric citation norms. While this hypothesis warrants further careful investigation, it suggests the complex interplay between author prominence and citation behavior.
Finally, the local coauthorship network of the first and last authors is linked to the over/under-citation made by their papers (Fig.~\ref{fig:6}(d)), which is consistent with the previous result in neuroscience \cite{dworkin2020}. 
In particular, MM papers whose male authors have a high proportion of male coauthors (i.e., top half of the \(\text{MA}_{\text{or}}\) range) display a greater absolute magnitude of over/under-citation than those whose male authors have fewer male coauthors (i.e., bottom half of the \(\text{MA}_{\text{or}}\) range).
WW papers with higher male representation in their coauthorship networks (i.e., top half of the \(\text{MA}_{\text{or}}\) range) exhibit stronger over-citation of WW papers. This result may indicate that women in male-dominated coauthorship networks could be more acutely aware of the gendered citation imbalance and have citation norms intended to address it.

\section{Discussion}

In this study, we investigated the extent of gendered citation imbalance in computer science. 
We found that papers with women as the first or last authors receive fewer citations across various subfields, topics, and venues in computer science. 
Previous studies analyzed gender imbalance in researcher demographics, productivity, research careers, and collaboration patterns in computer science \cite{way2016, jadidi2018, huang2020, laberge2022, morgan2021, lietz2024, holman2018, llwang2021, hajibabaei2022}. 
Our study extends previous findings on the gender imbalance in computer science in terms of citation practices. 
Our results indicate citation practices in the unique conference-oriented culture of computer science may reinforce the gendered citation imbalance.

We found that conference papers in computer science are more prone to gendered citation imbalances--both in making and receiving citations--than journal papers. 
Several factors may contribute to this disparity. 
One is the highly compressed peer-review cycle at computer science conferences, often restricted to a few weeks.
While these short turnaround times expedite decisions and knowledge dissemination, they raise concerns about review depth and quality.
Vardi highlighted the ``extreme time and workload pressures'' experienced by conference program committees \cite{vardi2009}. 
Fortnow argued that deadline-driven conference culture can lead to arbitrary decisions and prioritize speed over thorough vetting \cite{vardi2009, fortnow2009}. 
These abbreviated timelines and rushed evaluations may inadvertently foster selective citation practices that exacerbate gendered citation imbalance, although further quantitative investigation is needed to determine the precise extent of this effect.

Our findings indicate that gendered citation imbalance is particularly pronounced in top-tier conference papers in computer science. 
In addition to conventional fixed schedules and brief peer-review periods, the rapid expansion of these conferences has surpassed the capacity of program committees, significantly increasing reviewers' workloads \cite{vardi2009, fortnow2009}. 
As early as 2009, Birman and Schneider reported that top-tier computer systems conferences were overwhelmed by submissions, leading to overworked reviewers and potentially cursory paper evaluations \cite{birman2009}. 
In 2019, an expert panel from the flagship conference on Computer and Communications Security (the so-called CCS conference) reported a surge to 3,039 submissions across four top-tier computer security conferences in 2020, raising concerns about the ability of reviewers to adequately assess such a large volume of manuscripts \cite{soneji2022}.
In response, several top-tier conferences have begun offering rebuttal opportunities to authors \cite{huang2023}. 
Future studies should examine how these peer-review dynamics at top-tier conferences shape citation practices, ensuring fair scholarly recognition in computer science.

Computer science conferences have traditionally enforced strict page limits to standardize submissions and facilitate brief review cycles \cite{halpern2011}. 
While these limits remain common, many prestigious conferences now allow longer, more content-rich papers. 
For instance, the flagship conference in human-computer interaction (the so-called CHI conference) lifted formal page limits in 2016, spurring increases in both paper length and references \cite{oppenlaender2024}. 
Similarly, Geiger's analysis of another top-tier conference showed that once page limits were removed, average paper length rose, reflecting community norms favoring more thorough reporting \cite{geiger2019}. 
These shifts try to balance efficiency for short review periods, physical constraints on conference proceedings, and the goal of more comprehensive research reporting.  
Although evolving page-limit policies appear to boost reference counts, it remains unclear how this trend will influence gendered citation imbalance in computer science.

The pronounced difference between conferences and journals in terms of gendered citation imbalance may also be partially driven by ``topic proximity'' in citation practices. 
Recent work in physics has shown that gendered citation imbalance is often amplified when authors cite work they are less familiar with (i.e., low topic proximity) \cite{teich2022}.
A possible hypothesis in computer science is that the constraints of conference publications (e.g., tight deadlines and page limits) may lead authors to prioritize highly visible or cited papers when referencing work outside their immediate expertise. 
Because computer science has historically been male-dominated, as we showed, this tendency could amplify gendered citation imbalances relative to journal papers, where authors may have more resources to conduct comprehensive literature searches.
Future work could investigate this possible association by measuring topic proximity using paper classifications or text embeddings.

Previous studies deployed reference models to characterize citation behaviors. 
Uzzi et al.~investigated atypical combinations of citations associated with the impact of a paper by randomizing citations using a Monte Carlo algorithm \cite{uzzi2013}.
Kojaku et al.~examined anomalous citation patterns across journals using a reference model that accounts for scientific communities and journal size \cite{kojaku2021}.
These reference models are not intended to quantify imbalances in citations received by papers because they preserve the number of citations received by each paper in the original network.
Dworkin et al.~introduced a reference model for quantifying the gender imbalance in citations received by papers \cite{dworkin2020}.
The model is not designed to investigate the associations between network properties and gendered citation imbalance because it preserves only the number of citations made by each paper in the original network. 
To address these limitations, we developed a family of reference models that preserve the number of citations made by each paper, along with two network properties: homophily in citations and heterogeneity in the number of citations received per paper.

Using reference models for citation networks, we found that homophily in citations strongly contributes to gendered citation imbalance. 
Previous studies have also focused on homophily in citations regarding authors' gender \cite{tekles2022, zhou2024}. 
Tekles et al. reported that in biological and medical fields, gendered citation patterns are largely driven by homophily regarding the research topic \cite{tekles2022}. 
Zhou et al. found that author's gender homophily partially explains citation behavior in life sciences \cite{zhou2024}.
By controlling for homophily in citations regarding country of affiliation, research topic, and venue rank, we identified its contribution to the gendered citation imbalance. 
Together with these previous studies, our findings highlight the significance of controlling for citation homophily when analyzing gendered citation patterns.

By conducting a matched-pair analysis controlling for key paper-level characteristics, we found that the prominence of the first or last author and their local coauthorship network contribute to gender imbalance in citations made by papers. 
This observation aligns with previous findings that these factors partially shape researchers' practices. 
Indeed, at top computer science conferences, manuscripts with prominent authors had higher scores than those without in peer reviews where the reviewers know the authors' names \cite{tomkins2017, sun2022}.
Similarly, local coauthorship network influences gender-based citation patterns of neuroscience and physics researchers \cite{dworkin2020, teich2022}. 
Using Yates's chi-squared test \cite{yates1934}, we found that these characteristics are not independent (see Supplementary Section S7). 
The interaction between these factors may amplify gendered citation imbalance, warranting systematic investigation into their combined effects.

Our study has several limitations. 
First, the gender assignment method we employed underperformed for Chinese-name benchmark data \cite{nakajima2023}.
As a result, we failed to assign gender categories to many papers written by Chinese authors.
Given the rapid increase in China's contribution to computer science research \cite{gomez2022}, future work should examine gendered citation imbalances in papers authored by Chinese researchers. 
Second, although we assumed that the first and last authors played leading roles in each paper, this assumption may not always hold. 
Indeed, gender imbalance can emerge in how credit is assigned among multiple authors \cite{ross2022}, and some computer science subfields adopt alphabetical ordering by surname \cite{solomon2009}. 
Further work is needed to implement gender-aware approaches to author credit and assess gendered citation imbalance in these subfields. 
Third, additional citation imbalances may exist regarding the author's country of affiliation \cite{gomez2022}, nationality \cite{nakajima2023}, and race \cite{kozlowski2022}. 
Although we focused on citation imbalance regarding the author's gender, our quantitative framework is not restricted to that context. 
For instance, citation imbalances likely exist between research-intensive and other countries \cite{gomez2022}, which could be analyzed using reference models controlling for network properties and paper characteristics other than country of affiliation. 
Systematic investigations into citation imbalances across various aspects of authors' identities are an important direction for future research.
Fourth, we acknowledge that computer science is becoming increasingly interdisciplinary \cite{chakraborty2018}. 
This trend makes it challenging to exhaustively define the field's research landscape, and our analysis might omit certain relevant publications, conferences, or journals in interdisciplinary areas where computer science plays a primary role.
Fifth, our analysis treats venue rank, author prominence, and co-authorship networks as static attributes, although they are time-varying. A longitudinal analysis incorporating these time-varying attributes warrants further investigation into how the evolution of an author's or venue's prestige over time may influence citation practices.

Despite these limitations, our study provides a systematic and large-scale analysis of gendered citation imbalance in computer science, encompassing both conferences and journals. 
Our findings are expected to provide actionable insights to inform ongoing efforts toward fostering a more equitable, inclusive, and sustainable academic environment.

\begin{appendices}

\section{Comparison of reference models} \label{appendix:a}

\begin{figure}[p]
\centering
\includegraphics[width=0.99\textwidth]{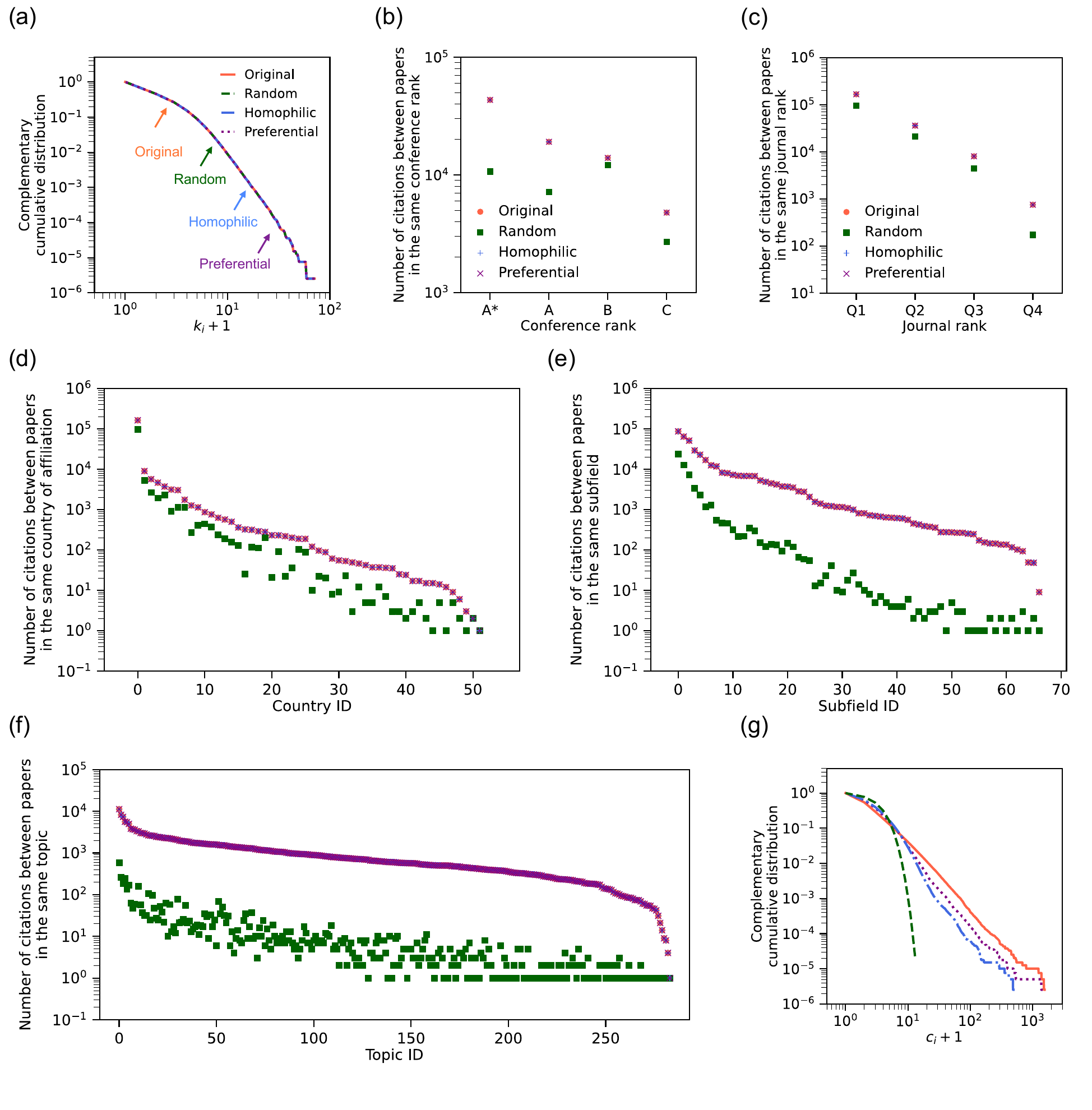}
\caption{
Comparison of structural properties between the original network and the reference models.
(a) Distribution of the number of citations made by each paper.  
(b)--(f) Homophilic citation patterns with respect to (b) conference rank, (c) journal rank, (d) affiliation country, (e) subfield, and (f) topic.  
(g) Distribution of the number of citations received by each paper.  
`Original' denotes the original network, `Random' the random-draws model, `Homophilic' the homophilic-draws model, and `Preferential' the preferential-draws model.  
In (d)--(f), we focus on the 52 countries, 67 subfields, and 284 topics that exhibit at least one homophilic citation in both the original network and each reference model.  
Curves that completely or substantially overlap are indicated by arrows and labels.
}
\label{fig:a1}
\end{figure}

We compare structural properties of the original network with those of randomized networks generated by our three reference models. 
First, all models exactly preserve the distribution of citations made by each paper (Fig.~\ref{fig:a1}(a)). 
Second, the random-draws model destroys homophilic citation patterns (in conference rank, journal rank, country of affiliation, and research topic), whereas the homophilic-draws and preferential-draws models preserve these patterns (Figs.~\ref{fig:a1}(b)--(f)). 
Third, the random-draws model poorly preserves the original heterogeneity in the number of citations received per paper, while the homophilic-draws and preferential-draws models approximate it more closely, with the latter being closer (Fig.~\ref{fig:a1}(g)). These outcomes align with our expectations.

\end{appendices}

\section*{Acknowledgements}

We thank Naoki Masuda (State University of New York at Buffalo) for his feedback. 
This work was supported by Japan Science and Technology Agency (JST) as part of Adopting Sustainable Partnerships for Innovative Research Ecosystem (ASPIRE), Grant Number JPMJAP2328. 
KN thanks the financial support by JSPS KAKENHI Grant Number 24K21056. 
YS thanks the support by JST Presto Grant Number JPMJPR21C5.

\section*{Competing interests}

The authors declare that they have no competing interests.

\section*{Data availability}

The OpenAlex snapshot from September 2024 is available at \url{https://docs.openalex.org/download-all-data/openalex-snapshot}.  
The DBLP snapshot from July 2024 is available at \url{https://blog.dblp.org/tag/snapshot/}.  
The CORE ranking data is available at \url{https://portal.core.edu.au/conf-ranks/}.  
The SCImago Journal Rank data is available at \url{https://www.scimagojr.com/journalrank.php?area=1700&year=2021}.  
The Gender API is available for a fee at \url{https://gender-api.com/}. 
Data from the 3,720,575 papers integrated between OpenAlex and DBLP (see Section \ref{section:2.1.3}) is available at \url{https://doi.org/10.5281/zenodo.16537950}.

\newpage

\begin{center}
\vspace*{12pt}
{\Large Supplementary Information for:\\
\vspace{12pt}
Systemic Gendered Citation Imbalance in Computer Science: Evidence from Conferences and Journals
}
\vspace{12pt} \\
\end{center}

\setcounter{figure}{0}
\setcounter{table}{0}
\setcounter{section}{0}

\renewcommand{\thesection}{S\arabic{section}}
\renewcommand{\thefigure}{S\arabic{figure}}
\renewcommand{\thetable}{S\arabic{table}}
\renewcommand{\theequation}{S\arabic{equation}}
\renewcommand{\thealgorithm}{S\arabic{algorithm}}

\begin{center}
\author{Kazuki Nakajima, Yuya Sasaki, Sohei Tokuno, and George Fletcher}
\vspace{24pt} \\
\end{center}

\section{Additional information on our dataset}

\subsection{Flow diagram of the filtering process for papers}

\begin{figure}[t]
\centering
\includegraphics[width=1\textwidth]{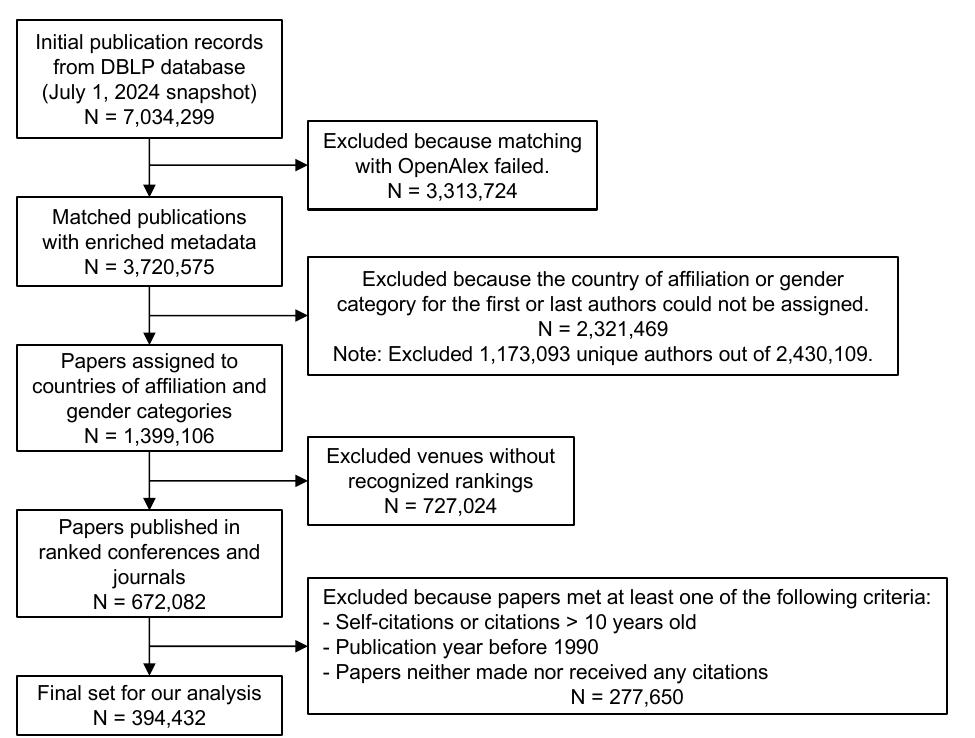}
\caption{
Flow diagram of the paper filtering process.
The diagram illustrates the step-by-step reduction from the initial publication records in the DBLP database to the final set of papers for our analysis.
At each step, the reason and the number of excluded papers are shown.
We also note the number of unique authors excluded because their country of affiliation or gender could not be assigned.
}
\label{fig:s1}
\end{figure}

Figure~\ref{fig:s1} shows a flow diagram of the filtering process for papers.

\subsection{Analysis of the representativeness of our final dataset}

To check the representativeness of our final dataset, we tracked the proportions of the four gender categories (i.e., ``MM'', ``MW'', ``WM'', and ``WW'') through the filtering process for papers.
We found that the relative representation of each category remained stable throughout the process:
\begin{itemize}
\item \textbf{Papers assigned to countries of affiliation and gender categories ($N = 1,399,106$):} MM: 75.2\%, MW: 7.4\%, WM: 10.7\%, WW: 6.6\%
\item \textbf{Papers published in ranked conferences and journals ($N = 672,082$):} MM: 76.0\%, MW: 7.3\%, WM: 10.5\%, WW: 6.1\%
\item \textbf{Final set for our analysis ($N = 394,432$):} MM: 75.6\%, MW: 7.8\%, WM: 11.1\%, WW: 5.5\%
\end{itemize}

We also compare the distribution of venue types and subfields between the initial matched dataset, $\mathcal{D}$, and the final dataset. 
The proportion of conference papers decreased in the final dataset (from 51.8\% to 39.3\%), while the proportion of journal papers increased. 
This shift reflects our quality criterion of including only venues ranked by CORE or SCImago; more unranked conferences than journals were excluded from the initial matched dataset.
Regarding subfields, the filtering process increased the relative share of the 11 computer science subfields while reducing the share of other subfields (from 51.9\% to 41.6\%; see Table~\ref{table:s1}). 
This indicates that our filtering concentrated the dataset on the central research areas of the discipline.

\subsection{Demographic comparison of included and excluded authors}

Table~\ref{table:s2} compares the distributions of the country of affiliation between the set of authors assigned a country of affiliation and the set of authors assigned both a country of affiliation and gender. 
We list individual countries until their cumulative share of authors assigned a country of affiliation reaches 90\%, with the remaining countries grouped as ``Others''. 
The most prominent shift is the sharp decrease in the relative share of authors from China. 
Similar decreases are also observed for authors from Taiwan, South Korea, Malaysia, Singapore, and Thailand. 
These shifts reflect the higher gender ambiguity associated with names from these regions, which often failed to meet our gender inference accuracy thresholds.

\subsection{Summary of metadata for papers}
Table~\ref{table:s3} summarizes the metadata of the papers included in our dataset.

\subsection{Additional statistics}
\begin{figure}[t]
\centering
\includegraphics[width=1\textwidth]{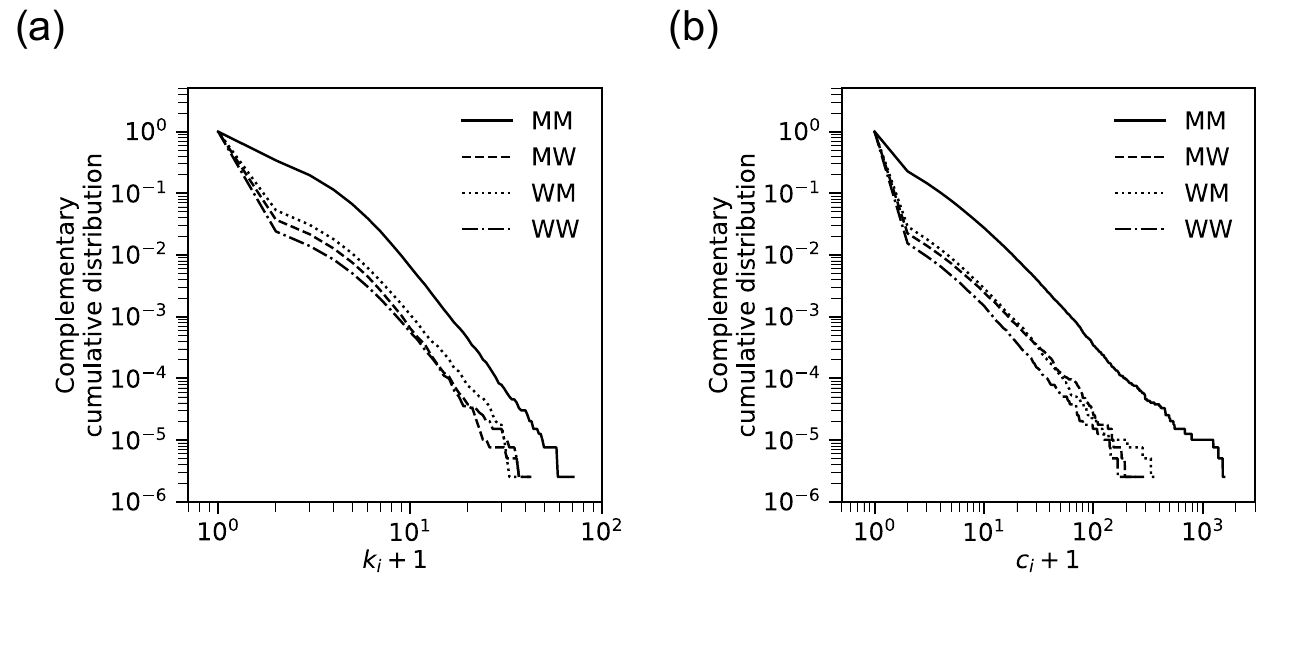}
\caption{
Survival functions of citation counts for each gender category. (a) Number of citations made by a paper, denoted by $k_i$, plus one. (b) Number of citations received by a paper, denoted by $c_i$, plus one.
}
\label{fig:s2}
\end{figure}

Table~\ref{table:s4} shows the number of papers across each subfield-gender category combination. 
Table~\ref{table:s5} shows the number of papers across each venue rank-gender category combination. 
Figure~\ref{fig:s2} shows the survival functions of citations made and received by papers in each gender category.

\section{Pseudocode of the preferential-draws model}

Algorithm \ref{alg:s1} shows a pseudocode of the preferential-draws model.

\begin{algorithm}[h]
\caption{Preferential-draws model.}
\label{alg:s1}
\begin{algorithmic}[1]
\REQUIRE Citation network: $(V, E)$ and set of characteristics of the paper: $S$.
\ENSURE Randomized network: $(V, E_{\text{rand}})$
\STATE{Sort the papers by publication date in ascending order: $v_{x_1}, \ldots, v_{x_N}$.}
\STATE{Initialize $c_{j, l, \text{PD}}$ with zero for any $j=1, \ldots, N$ for any $l=1, \ldots, N$.}
\STATE{$E_{\text{rand}} \leftarrow$ an empty list.}
\FOR{$l=2, \ldots, N$}
\FOR{each citation $(v_{x_l}, v_{j}) \in E$}
\STATE{Compute the set $\overline{V}_{\text{HD}}(x_l, j, S)$.}
\STATE{$\overline{V}_{\text{PD}}(x_l, j, S) \leftarrow \{v_{j'}\ |\ v_{j'} \in \overline{V}_{\text{HD}}(x_l, j, S) \ \land \ \lfloor \ln (c_{j', l-1, \text{PD}} + 1) \rfloor = \lfloor \ln (c_{j, l-1, \text{PD}} + 1) \rfloor \}$.}
\STATE{Draw $v_{j'}$ uniformly at random from the set $\overline{V}_{\text{PD}}(x_l, j, S)$.}
\STATE{Append citation $(v_{x_l}, v_{j'})$ to $E_{\text{rand}}$.}
\FOR{$z=l, \ldots, N$}
\STATE{$c_{j', z, \text{PD}} \leftarrow c_{j', z, \text{PD}} + 1$.}
\ENDFOR
\ENDFOR
\ENDFOR
\RETURN{$(V, E_{\text{rand}})$}
\end{algorithmic}
\end{algorithm}

\section{Detailed results of the over/under-citation}

\begin{figure}[t]
\centering
\includegraphics[width=1\textwidth]{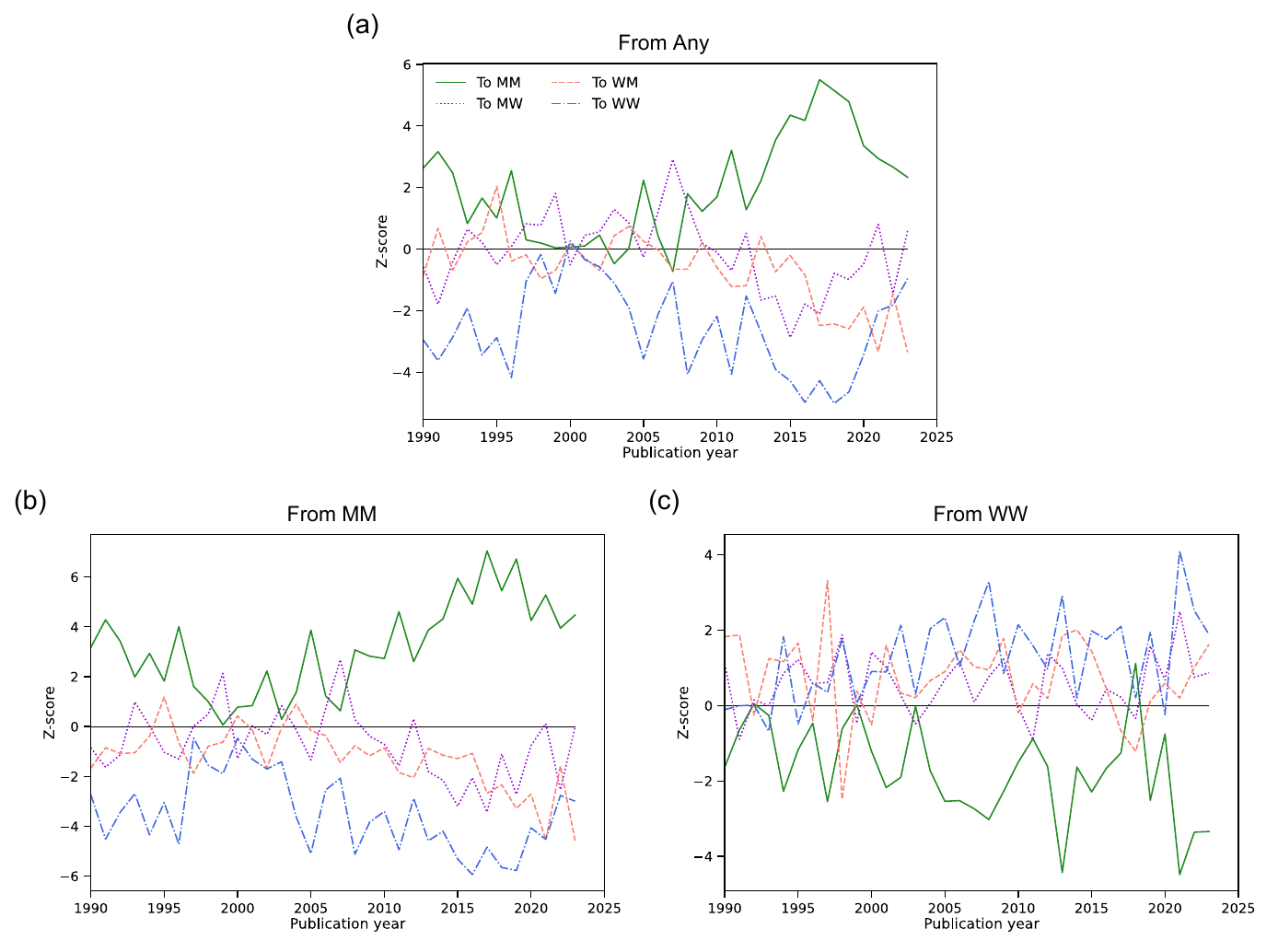}
\caption{
Temporal trends of the Z scores of the over/under-citation made by papers in computer science. (a): All papers. (b): MM papers. (c): WW papers.
}
\label{fig:s3}
\end{figure}

Tables~\ref{table:s6}-\ref{table:s8} show the over/under-citations under the random-draws, homophilic-draws, and preferential-draws models, respectively. 
Table~\ref{table:s9} shows the over/under-citation made by all papers to papers in a given gender category and in a given subfield. 
Table~\ref{table:s10} shows the over/under-citation made by all papers to papers in a given gender category and in a given venue type. 
Table~\ref{table:s11} shows the over/under-citation made by all papers to papers in a given gender category and in a given venue rank.
Figure~\ref{fig:s3} shows the temporal trends of the Z-scores corresponding to the over/under-citation values shown in Fig.~4 of the main text.

\section{Effects of self-citations on the over/under-citation}

We constructed a citation network with self-citations intact, comprising 418,090 papers and 820,258 citations. 
Recall that a citation is defined as a self-citation if there is any overlap in authorship between the citing paper \(u\) and the cited paper \(v\). 
We then calculated over/under-citations in this network. Table~\ref{table:s12} compares these values with those computed from the network in which self-citations were excluded.

\section{Effects of isolated papers on the over/under-citation}

We constructed a citation network with isolated papers (i.e., papers that neither make nor receive any citations) intact, comprising 542,550 papers and 752,742 citations. 
We then calculated over/under-citations in this network. Table~\ref{table:s13} compares these values with those computed from the network in which isolated papers were excluded.

\section{Detailed results of the matching experiments}

Tables \ref{table:s14}--\ref{table:s17} show the results of the matching experiments with respect to the author's gender (MM or WW), venue type, involvement of prominent authors, and local coauthorship network, respectively.

\section{Relationships between the three characteristics of the paper}

We investigate the relationships among our three characteristics: venue type, involvement of prominent authors, and local coauthorship network. 
Specifically, we use Yates's chi-squared test \cite{yates1934} for MM and WW papers separately. 
Fix a gender category \(g \in \{\text{MM}, \text{WW}\}\) and two characteristics, \(X\) and \(Y\). 
Each characteristic is a categorical variable with two possible categories for each paper. 
We denote by \(O_{g,i,j}\) the number of papers in gender category \(g\) that fall into category \(i\) for \(X\) (\(i=1,2\)) and category \(j\) for \(Y\) (\(j=1,2\)). 
Let $N_g = \sum_{i=1}^2 \sum_{j=1}^2 O_{g,i,j} $ be the total number of papers in category \(g\).
We define the test statistic by
\begin{align}
\chi^2 = \sum_{i=1}^2 \sum_{j=1}^2 (|O_{g, i, j} - E_{g, i, j}| - 0.5)^2 / E_{g, i, j},
\label{eq:5}
\end{align}
where
\begin{align}
E_{g, i, j} &= N_g p_{g, i} q_{g, j}, \label{eq:6} \\ 
p_{g, i} &= \sum_{j=1}^2 \frac{O_{g, i, j}}{N_g}, \label{eq:7} \\
q_{g, j} &= \sum_{i=1}^2 \frac{O_{g, i, j}}{N_g}.
\label{eq:8}
\end{align}
The null hypothesis is that $X$ is independent of $Y$ for the papers in $g$.
If the $p$ value of the test is less than 0.001, we reject the null hypothesis.

Table \ref{table:s18} shows the test statistic and \(p\)-value for each combination \((g, X, Y)\). 
For MM papers, every pair of characteristics is not independent. 
For WW papers, every characteristic pair except ``venue type versus local coauthorship network'' is not independent.


\newpage

\newpage

\newpage

\newpage

\begin{table}[p]
\centering
\caption{Comparison of subfield distributions between the initial matched dataset and the final dataset. Subfields not belonging to the `Computer Science' field in OpenAlex are grouped as `Others'.}
\label{table:s1}
\begin{tabular}{lrr}
\hline
Subfield & Initial Set (\%) & Final Set (\%) \\
\hline
Artificial Intelligence & 12.83 & 16.50 \\
Computational Theory and Mathematics & 3.53 & 3.86 \\
Computer Graphics and Computer-Aided Design & 0.72 & 1.17 \\
Computer Networks and Communications & 9.76 & 11.63 \\
Computer Science Applications & 0.87 & 1.35 \\
Computer Vision and Pattern Recognition & 8.25 & 9.04 \\
Hardware and Architecture & 2.34 & 3.19 \\
Human-Computer Interaction & 1.10 & 1.71 \\
Information Systems & 5.13 & 6.18 \\
Signal Processing & 2.78 & 2.66 \\
Software & 0.78 & 1.13 \\
Others & 51.92 & 41.59 \\
\hline
\end{tabular}
\end{table}

\newpage

\begin{table}[h]
\centering
\caption{
Comparison of the distributions of country of affiliation between authors assigned a country ($p_1$) and authors assigned both a country and a gender ($p_2$). Countries are listed individually until the cumulative share of $p_1$ reaches 90\%, with the remaining countries grouped as `Others'.}
\label{table:s2}
\begin{tabular}{lrr}
\hline
Country & $p_1$ (\%) & $p_2$ (\%) \\
\hline
United States & 22.21 & 26.52 \\
China & 16.22 & 1.77 \\
Germany & 5.71 & 8.07 \\
Japan & 4.98 & 6.00 \\
India & 4.25 & 4.31 \\
United Kingdom & 4.22 & 5.20 \\
South Korea & 3.41 & 2.30 \\
Italy & 3.15 & 4.10 \\
Taiwan & 2.95 & 0.56 \\
Canada & 2.86 & 3.46 \\
France & 2.67 & 3.55 \\
Brazil & 2.21 & 3.19 \\
Australia & 1.73 & 2.04 \\
Netherlands & 1.50 & 1.76 \\
Iran & 1.26 & 1.73 \\
Spain & 1.20 & 1.65 \\
Switzerland & 0.96 & 1.30 \\
Greece & 0.84 & 1.13 \\
Sweden & 0.81 & 1.13 \\
Russia & 0.81 & 0.90 \\
Malaysia & 0.79 & 0.68 \\
Belgium & 0.73 & 0.99 \\
Austria & 0.70 & 1.02 \\
Portugal & 0.70 & 1.02 \\
Finland & 0.69 & 0.88 \\
Singapore & 0.68 & 0.44 \\
Israel & 0.66 & 0.73 \\
Pakistan & 0.58 & 0.79 \\
Thailand & 0.49 & 0.37 \\
Denmark & 0.48 & 0.67 \\
Others & 9.55 & 11.74 \\
\hline
\end{tabular}
\end{table}

\begin{table}[t]
 \begin{center}
   \caption{Summary of paper metadata.}
   \label{table:s3}
\begin{tabular}{|l|l|} \hline
Attribute & Data source \\ \hline \hline
Title & OpenAlex \\
Publication date & OpenAlex \\
Primary research topic & OpenAlex \\
Primary research subfield & OpenAlex \\
Authors' names & OpenAlex \\
Authors' affiliations & OpenAlex \\
Authors' IDs & DBLP \\
Authorship order & DBLP \\
Papers cited by the paper & OpenAlex \\
Venue type (i.e., conference or journal) & DBLP \\
Venue name & DBLP \\
Venue rank & DBLP, CORE, and SCImago \\
Country of affiliation & OpenAlex and DBLP \\
Gender category & OpenAlex, DBLP, Gender API \\
\hline
 \end{tabular}
 \end{center}
\end{table}

\begin{table}[t]
   \caption{Number of papers by subfield and gender category. }
   \label{table:s4}
 \begin{center}
\begin{tabular}{|l|c|c|c|c|} \hline 
Subfield & MM & MW & WM & WW \\ \hline \hline
Artificial Intelligence & 49,852 & 5,156 & 6,653 & 3,413\\
 & (76.7\%) & (7.9\%) & (10.2\%) & (5.2\%)\\ \hline
Computational Theory and Mathematics & 12,383 & 938 & 1,224 & 698\\
 & (81.2\%) & (6.2\%) & (8.0\%) & (4.6\%)\\ \hline
Computer Graphics and Computer-Aided Design & 4,039 & 205 & 261 & 92\\
 & (87.8\%) & (4.5\%) & (5.7\%) & (2.0\%)\\ \hline
Computer Networks and Communications & 36,979 & 3,102 & 4,425 & 1,362\\
 & (80.5\%) & (6.8\%) & (9.7\%) & (3.0\%)\\ \hline
Computer Science Applications & 3,107 & 571 & 866 & 764\\
 & (58.5\%) & (10.8\%) & (16.3\%) & (14.4\%)\\ \hline
Computer Vision and Pattern Recognition & 28,040 & 2,552 & 3,918 & 1,135\\
 & (78.6\%) & (7.2\%) & (11.0\%) & (3.2\%)\\ \hline
Hardware and Architecture & 10,790 & 733 & 851 & 227\\
 & (85.6\%) & (5.8\%) & (6.8\%) & (1.8\%)\\ \hline
Human-Computer Interaction & 4,168 & 730 & 1,085 & 743\\
 & (62.0\%) & (10.8\%) & (16.1\%) & (11.1\%)\\ \hline
Information Systems & 17,497 & 2,091 & 3,081 & 1,725\\
 & (71.7\%) & (8.6\%) & (12.6\%) & (7.1\%)\\ \hline
Signal Processing & 8,266 & 750 & 1,136 & 323\\
 & (78.9\%) & (7.2\%) & (10.8\%) & (3.1\%)\\ \hline
Software & 3,321 & 408 & 461 & 265\\
 & (74.5\%) & (9.2\%) & (10.3\%) & (6.0\%)\\ \hline
 \end{tabular}
 \end{center}
\end{table}

\begin{table}[t]
   \caption{Number of papers by venue rank and gender category.}
   \label{table:s5}
 \begin{center}
\begin{tabular}{|l|c|c|c|c|} \hline
Rank & MM & MW & WM & WW \\ \hline \hline
$\text{A}^*$ & 32,325 & 3,062 & 3,692 & 1,680\\
 & (79.3\%) & (7.5\%) & (9.1\%) & (4.1\%)\\ \hline
A & 28,020 & 3,064 & 3,851 & 2,045\\
 & (75.8\%) & (8.3\%) & (10.4\%) & (5.5\%)\\ \hline
B & 39,782 & 4,283 & 5,943 & 2,091\\
 & (76.4\%) & (8.2\%) & (11.4\%) & (4.0\%)\\ \hline
C & 18,261 & 2,101 & 3,149 & 1,475\\
 & (73.1\%) & (8.4\%) & (12.6\%) & (5.9\%)\\ \hline
Q1 & 102,895 & 10,381 & 15,173 & 7,068\\
 & (75.9\%) & (7.7\%) & (11.2\%) & (5.2\%)\\ \hline
Q2 & 49,132 & 5,509 & 8,010 & 4,962\\
 & (72.7\%) & (8.2\%) & (11.8\%) & (7.3\%)\\ \hline
Q3 & 23,105 & 2,134 & 3,239 & 1,909\\
 & (76.0\%) & (7.0\%) & (10.7\%) & (6.3\%)\\ \hline
Q4 & 4,543 & 436 & 670 & 442\\
 & (74.5\%) & (7.2\%) & (11.0\%) & (7.3\%)\\ \hline
 \end{tabular}
 \end{center}
\end{table}

\begin{table}[t]
   \caption{Over/under-citation derived from the random-draws model. We present the over/under-citation made by a set of papers (e.g., those in a given gender category or all papers) to another set of papers (also defined by gender category). Here, \(n_{\text{obs}}\) is the number of citations observed in the original network, while \(\mu\) and \(\sigma\) are the mean and standard deviation, respectively, of citation counts computed across 100 randomized networks.}
   \label{table:s6}
 \begin{center}
\begin{tabular}{|c|c|C|C|C|A|C|C|} \hline
   Made by & Received by & $n_{\text{obs}}$ & $\mu$ & $\sigma$ & Over/under-citation (\%) & Z score & $p$ value \\ \hline \hline
MM & MM & 454733 & 438991.5 & $304.9$ & $3.6$ & $51.63$ & $< 0.001$ \\
MM & MW & 38826 & 40627.3 & $179.8$ & $-4.4$ & $-10.02$ & $< 0.001$ \\
MM & WM & 47068 & 53171.2 & $234.7$ & $-11.5$ & $-26.01$ & $< 0.001$ \\
MM & WW & 21853 & 29689.9 & $170.9$ & $-26.4$ & $-45.86$ & $< 0.001$ \\ \hline
MW & MM & 45992 & 47472.8 & $106.1$ & $-3.1$ & $-13.96$ & $< 0.001$ \\
MW & MW & 5282 & 4594.3 & $64.7$ & $15.0$ & $10.64$ & $< 0.001$ \\
MW & WM & 6486 & 6157.7 & $72.9$ & $5.3$ & $4.50$ & $< 0.001$ \\
MW & WW & 3688 & 3223.2 & $47.2$ & $14.4$ & $9.85$ & $< 0.001$ \\ \hline
WM & MM & 64414 & 67369.8 & $131.7$ & $-4.4$ & $-22.44$ & $< 0.001$ \\
WM & MW & 7692 & 6777.6 & $89.4$ & $13.5$ & $10.23$ & $< 0.001$ \\
WM & WM & 10244 & 9189.0 & $94.3$ & $11.5$ & $11.19$ & $< 0.001$ \\
WM & WW & 5586 & 4599.6 & $65.3$ & $21.4$ & $15.10$ & $< 0.001$ \\ \hline
WW & MM & 27220 & 31674.4 & $86.4$ & $-14.1$ & $-51.54$ & $< 0.001$ \\
WW & MW & 3783 & 3007.8 & $55.2$ & $25.8$ & $14.06$ & $< 0.001$ \\
WW & WM & 5115 & 4021.0 & $62.6$ & $27.2$ & $17.48$ & $< 0.001$ \\
WW & WW & 4760 & 2174.8 & $41.4$ & $118.9$ & $62.46$ & $< 0.001$ \\ \hline
All & MM & 592359 & 585508.6 & $357.2$ & $1.2$ & $19.18$ & $< 0.001$ \\
All & MW & 55583 & 55006.9 & $217.8$ & $1.0$ & $2.65$ & $0.004$ \\
All & WM & 68913 & 72539.0 & $279.9$ & $-5.0$ & $-12.95$ & $< 0.001$ \\
All & WW & 35887 & 39687.6 & $202.0$ & $-9.6$ & $-18.82$ & $< 0.001$ \\ \hline
 \end{tabular}
 \end{center}
\end{table}

\begin{table}[t]
   \caption{Over/under-citation derived from the homophilic-draws model.}
   \label{table:s7}
 \begin{center}
\begin{tabular}{|c|c|C|C|C|A|C|C|} \hline
   Made by & Received by & $n_{\text{obs}}$ & $\mu$ & $\sigma$ & Over/under-citation (\%) & Z score & $p$ value \\ \hline \hline
MM & MM & 454733 & 447203.8 & $255.1$ & $1.7$ & $29.51$ & $< 0.001$ \\
MM & MW & 38826 & 40175.6 & $179.5$ & $-3.4$ & $-7.52$ & $< 0.001$ \\
MM & WM & 47068 & 50282.7 & $195.5$ & $-6.4$ & $-16.44$ & $< 0.001$ \\
MM & WW & 21853 & 24817.9 & $132.2$ & $-11.9$ & $-22.43$ & $< 0.001$ \\ \hline
MW & MM & 45992 & 45988.4 & $87.1$ & $0.0$ & $0.04$ & $0.483$ \\
MW & MW & 5282 & 5099.5 & $60.2$ & $3.6$ & $3.03$ & $0.001$ \\
MW & WM & 6486 & 6662.1 & $58.0$ & $-2.6$ & $-3.03$ & $0.001$ \\
MW & WW & 3688 & 3698.0 & $49.8$ & $-0.3$ & $-0.20$ & $0.420$ \\ \hline
WM & MM & 64414 & 64612.6 & $101.1$ & $-0.3$ & $-1.97$ & $0.025$ \\
WM & MW & 7692 & 7556.1 & $73.0$ & $1.8$ & $1.86$ & $0.031$ \\
WM & WM & 10244 & 10208.7 & $82.0$ & $0.3$ & $0.43$ & $0.334$ \\
WM & WW & 5586 & 5558.5 & $59.0$ & $0.5$ & $0.47$ & $0.321$ \\ \hline
WW & MM & 27220 & 28031.5 & $71.1$ & $-2.9$ & $-11.41$ & $< 0.001$ \\
WW & MW & 3783 & 3661.3 & $43.5$ & $3.3$ & $2.80$ & $0.003$ \\
WW & WM & 5115 & 4953.4 & $51.9$ & $3.3$ & $3.11$ & $< 0.001$ \\
WW & WW & 4760 & 4231.7 & $52.0$ & $12.5$ & $10.16$ & $< 0.001$ \\ \hline
All & MM & 592359 & 585836.3 & $324.6$ & $1.1$ & $20.10$ & $< 0.001$ \\
All & MW & 55583 & 56492.6 & $217.7$ & $-1.6$ & $-4.18$ & $< 0.001$ \\
All & WM & 68913 & 72106.9 & $242.3$ & $-4.4$ & $-13.18$ & $< 0.001$ \\
All & WW & 35887 & 38306.2 & $170.5$ & $-6.3$ & $-14.19$ & $< 0.001$ \\ \hline
 \end{tabular}
 \end{center}
\end{table}

\begin{table}[t]
   \caption{Over/under-citation derived from the preferential-draws model.}
   \label{table:s8}
 \begin{center}
\begin{tabular}{|c|c|C|C|C|A|C|C|} \hline
   Made by & Received by & $n_{\text{obs}}$ & $\mu$ & $\sigma$ & Over/under-citation (\%) & Z score & $p$ value \\ \hline \hline
MM & MM & 454733 & 449454.7 & $341.7$ & $1.2$ & $15.45$ & $< 0.001$ \\
MM & MW & 38826 & 39613.1 & $236.0$ & $-2.0$ & $-3.34$ & $< 0.001$ \\
MM & WM & 47068 & 48780.3 & $233.9$ & $-3.5$ & $-7.32$ & $< 0.001$ \\
MM & WW & 21853 & 24631.9 & $169.4$ & $-11.3$ & $-16.40$ & $< 0.001$ \\ \hline
MW & MM & 45992 & 46142.0 & $86.3$ & $-0.3$ & $-1.74$ & $0.041$ \\
MW & MW & 5282 & 5076.2 & $60.2$ & $4.1$ & $3.42$ & $< 0.001$ \\
MW & WM & 6486 & 6506.2 & $61.2$ & $-0.3$ & $-0.33$ & $0.371$ \\
MW & WW & 3688 & 3723.6 & $48.6$ & $-1.0$ & $-0.73$ & $0.232$ \\ \hline
WM & MM & 64414 & 64799.4 & $111.4$ & $-0.6$ & $-3.46$ & $< 0.001$ \\
WM & MW & 7692 & 7507.8 & $69.1$ & $2.5$ & $2.67$ & $0.004$ \\
WM & WM & 10244 & 10029.6 & $85.6$ & $2.1$ & $2.50$ & $0.006$ \\
WM & WW & 5586 & 5599.2 & $61.9$ & $-0.2$ & $-0.21$ & $0.416$ \\ \hline
WW & MM & 27220 & 27979.6 & $71.7$ & $-2.7$ & $-10.60$ & $< 0.001$ \\
WW & MW & 3783 & 3644.2 & $51.2$ & $3.8$ & $2.71$ & $0.003$ \\
WW & WM & 5115 & 4916.9 & $62.0$ & $4.0$ & $3.20$ & $< 0.001$ \\
WW & WW & 4760 & 4337.2 & $54.6$ & $9.7$ & $7.74$ & $< 0.001$ \\ \hline
All & MM & 592359 & 588375.6 & $427.6$ & $0.7$ & $9.32$ & $< 0.001$ \\
All & MW & 55583 & 55841.4 & $302.3$ & $-0.5$ & $-0.85$ & $0.196$ \\
All & WM & 68913 & 70233.1 & $288.2$ & $-1.9$ & $-4.58$ & $< 0.001$ \\
All & WW & 35887 & 38291.9 & $217.5$ & $-6.3$ & $-11.06$ & $< 0.001$ \\ \hline
 \end{tabular}
 \end{center}
\end{table}

\begin{table}[t]
   \caption{Over/under-citation by all papers to papers by a gender category and subfield.}
   \label{table:s9}
 \begin{center}
\begin{tabular}{|l|B|C|C|C|A|C|C|} \hline
   Subfield & Gender category & $n_{\text{obs}}$ & $\mu$ & $\sigma$ & Over/under-citation (\%) & Z score & $p$ value \\ \hline \hline
Artificial Intelligence & MM & 115843 & 113620.9 & $214.7$ & $2.0$ & $10.35$ & $< 0.001$\\
 & WW & 6324 & 7000.1 & $125.6$ & $-9.7$ & $-5.38$ & $< 0.001$ \\ \hline
Computational Theory & MM & 21814 & 21427.3 & $57.6$ & $1.8$ & $6.71$ & $< 0.001$\\
and Mathematics & WW & 951 & 1077.9 & $32.2$ & $-11.8$ & $-3.94$ & $< 0.001$ \\ \hline
Computer Graphics and & MM & 11607 & 11426.6 & $34.7$ & $1.6$ & $5.20$ & $< 0.001$\\
Computer-Aided Design & WW & 158 & 223.2 & $17.1$ & $-29.2$ & $-3.80$ & $< 0.001$ \\ \hline
Computer Networks & MM & 79513 & 80036.4 & $145.9$ & $-0.7$ & $-3.59$ & $< 0.001$\\
and Communications & WW & 2486 & 2653.5 & $73.6$ & $-6.3$ & $-2.28$ & $0.011$ \\ \hline
Computer Science & MM & 5980 & 6195.5 & $49.6$ & $-3.5$ & $-4.35$ & $< 0.001$\\
Applications & WW & 1407 & 1352.8 & $37.6$ & $4.0$ & $1.44$ & $0.075$ \\ \hline
Computer Vision and & MM & 69467 & 68487.4 & $176.9$ & $1.4$ & $5.54$ & $< 0.001$\\
Pattern Recognition & WW & 1787 & 2208.7 & $72.6$ & $-19.1$ & $-5.81$ & $< 0.001$ \\ \hline
Hardware and & MM & 24161 & 24200.2 & $55.3$ & $-0.2$ & $-0.71$ & $0.239$\\
Architecture & WW & 517 & 588.2 & $22.8$ & $-12.1$ & $-3.12$ & $< 0.001$ \\ \hline
Human-Computer & MM & 9489 & 9235.9 & $56.5$ & $2.7$ & $4.48$ & $< 0.001$\\
Interaction & WW & 1412 & 1535.1 & $47.4$ & $-8.0$ & $-2.60$ & $0.005$ \\ \hline
Information Systems & MM & 39409 & 39386.1 & $104.3$ & $0.1$ & $0.22$ & $0.413$\\
 & WW & 3297 & 3539.7 & $61.9$ & $-6.9$ & $-3.92$ & $< 0.001$ \\ \hline
Signal Processing & MM & 14525 & 14405.2 & $55.9$ & $0.8$ & $2.14$ & $0.016$\\
 & WW & 376 & 422.3 & $19.5$ & $-11.0$ & $-2.37$ & $0.009$ \\ \hline
Software & MM & 7318 & 7336.6 & $48.5$ & $-0.3$ & $-0.38$ & $0.351$\\
 & WW & 464 & 677.9 & $25.0$ & $-31.6$ & $-8.56$ & $< 0.001$ \\ \hline
 \end{tabular}
 \end{center}
\end{table}

\begin{table}[t]
   \caption{Over/under-citation by all papers to papers by a gender category and venue type.}
   \label{table:s10}
 \begin{center}
\begin{tabular}{|l|B|C|C|C|A|C|C|} \hline
   Type & Gender category & $n_{\text{obs}}$ & $\mu$ & $\sigma$ & Over/under-citation (\%) & Z score & $p$ value \\ \hline \hline
Conference & MM & 221681 & 219533.3 & $226.9$ & $1.0$ & $9.47$ & $< 0.001$ \\
 & WW & 11520 & 13012.1 & $117.6$ & $-11.5$ & $-12.69$ & $< 0.001$ \\ \hline
Journal & MM & 370678 & 368820.3 & $316.5$ & $0.5$ & $5.87$ & $< 0.001$ \\
 & WW & 24367 & 25325.7 & $148.2$ & $-3.8$ & $-6.47$ & $< 0.001$ \\ \hline
 \end{tabular}
 \end{center}
\end{table}

\begin{table}[t]
   \caption{Over/under-citation by all papers to papers by a gender category and subfield.}
   \label{table:s11}
 \begin{center}
\begin{tabular}{|l|B|C|C|C|A|C|C|} \hline
   Rank & Gender category & $n_{\text{obs}}$ & $\mu$ & $\sigma$ & Over/under-citation (\%) & Z score & $p$ value \\ \hline \hline
A* & MM & 130750 & 129819.0 & $252.5$ & $0.7$ & $3.69$ & $< 0.001$\\
 & WW & 4391 & 5360.3 & $97.8$ & $-18.1$ & $-9.91$ & $< 0.001$ \\ \hline
A & MM & 55092 & 54469.0 & $150.5$ & $1.1$ & $4.14$ & $< 0.001$\\
 & WW & 3667 & 3759.5 & $82.0$ & $-2.5$ & $-1.13$ & $0.130$ \\ \hline
B & MM & 37248 & 36934.5 & $82.7$ & $0.8$ & $3.79$ & $< 0.001$\\
 & WW & 1633 & 1814.0 & $33.8$ & $-10.0$ & $-5.35$ & $< 0.001$ \\ \hline
C & MM & 14350 & 14169.3 & $54.6$ & $1.3$ & $3.31$ & $< 0.001$\\
 & WW & 1163 & 1184.6 & $28.0$ & $-1.8$ & $-0.77$ & $0.220$ \\ \hline
Q1 & MM & 257718 & 256007.4 & $263.3$ & $0.7$ & $6.50$ & $< 0.001$\\
 & WW & 16138 & 17130.2 & $153.6$ & $-5.8$ & $-6.46$ & $< 0.001$ \\ \hline
Q2 & MM & 70613 & 70598.2 & $125.3$ & $0.0$ & $0.12$ & $0.453$\\
 & WW & 6768 & 6748.6 & $83.5$ & $0.3$ & $0.23$ & $0.408$ \\ \hline
Q3 & MM & 23349 & 23139.9 & $61.1$ & $0.9$ & $3.42$ & $< 0.001$\\
 & WW & 1827 & 1993.2 & $38.4$ & $-8.3$ & $-4.33$ & $< 0.001$ \\ \hline
Q4 & MM & 3239 & 3238.2 & $16.6$ & $0.0$ & $0.05$ & $0.481$\\
 & WW & 300 & 301.6 & $9.1$ & $-0.5$ & $-0.18$ & $0.430$ \\ \hline
 \end{tabular}
 \end{center}
\end{table}

\begin{table}[t]
   \caption{Comparison of over/under-citations in networks with and without self-citations.}
   \label{table:s12}
 \begin{center}
\begin{tabular}{|c|c|A|C|C|A|C|C|} \cline{3-8}
   \multicolumn{1}{c}{} & \multicolumn{1}{c|}{} & \multicolumn{3}{c|}{With self-citations} & \multicolumn{3}{c|}{Without self-citations} \\ \hline
   Made by & Received by & Over/under-citation (\%) & Z score & $p$ value & Over/under-citation (\%) & Z score & $p$ value \\ \hline \hline
MM & MM & $2.1$ & $33.14$ & $< 0.001$ & $1.2$ & $14.47$ & $< 0.001$ \\
MM & MW & $-6.2$ & $-12.22$ & $< 0.001$ & $-2.0$ & $-3.48$ & $< 0.001$ \\
MM & WM & $-7.4$ & $-17.19$ & $< 0.001$ & $-3.5$ & $-7.03$ & $< 0.001$ \\
MM & WW & $-15.8$ & $-22.24$ & $< 0.001$ & $-11.4$ & $-17.10$ & $< 0.001$ \\ \hline
MW & MM & $-3.3$ & $-16.06$ & $< 0.001$ & $-0.3$ & $-1.68$ & $0.046$ \\
MW & MW & $24.7$ & $22.34$ & $< 0.001$ & $4.2$ & $3.82$ & $< 0.001$ \\
MW & WM & $-0.6$ & $-0.58$ & $0.280$ & $-0.2$ & $-0.18$ & $0.429$ \\
MW & WW & $2.0$ & $1.62$ & $0.053$ & $-1.5$ & $-1.25$ & $0.105$ \\ \hline
WM & MM & $-3.1$ & $-17.21$ & $< 0.001$ & $-0.6$ & $-3.18$ & $< 0.001$ \\
WM & MW & $2.7$ & $2.79$ & $0.003$ & $2.1$ & $2.05$ & $0.020$ \\
WM & WM & $15.1$ & $20.71$ & $< 0.001$ & $2.2$ & $2.45$ & $0.007$ \\
WM & WW & $2.2$ & $2.10$ & $0.018$ & $-0.2$ & $-0.16$ & $0.438$ \\ \hline
WW & MM & $-6.4$ & $-23.10$ & $< 0.001$ & $-2.7$ & $-11.76$ & $< 0.001$ \\
WW & MW & $0.2$ & $0.15$ & $0.442$ & $3.8$ & $3.14$ & $< 0.001$ \\
WW & WM & $1.1$ & $0.90$ & $0.183$ & $4.2$ & $4.00$ & $< 0.001$ \\
WW & WW & $31.1$ & $33.38$ & $< 0.001$ & $9.6$ & $8.65$ & $< 0.001$ \\ \hline
All & MM & $0.8$ & $11.50$ & $< 0.001$ & $0.7$ & $8.90$ & $< 0.001$ \\
All & MW & $-1.4$ & $-2.90$ & $0.002$ & $-0.5$ & $-1.02$ & $0.153$ \\
All & WM & $-2.7$ & $-6.97$ & $< 0.001$ & $-1.8$ & $-4.33$ & $< 0.001$ \\
All & WW & $-5.0$ & $-8.51$ & $< 0.001$ & $-6.4$ & $-11.12$ & $< 0.001$ \\ \hline
 \end{tabular}
 \end{center}
\end{table}

\begin{table}[t]
   \caption{Comparison of over/under-citations in networks with and without isolated papers.}
   \label{table:s13}
 \begin{center}
\begin{tabular}{|c|c|A|C|C|A|C|C|} \cline{3-8}
   \multicolumn{1}{c}{} & \multicolumn{1}{c|}{} & \multicolumn{3}{c|}{With isolated papers} & \multicolumn{3}{c|}{Without isolated papers} \\ \hline
   Made by & Received by & Over/under-citation (\%) & Z score & $p$ value & Over/under-citation (\%) & Z score & $p$ value \\ \hline \hline
MM & MM & $1.2$ & $15.45$ & $< 0.001$ & $1.2$ & $14.47$ & $< 0.001$ \\
MM & MW & $-1.4$ & $-2.58$ & $0.005$ & $-2.0$ & $-3.48$ & $< 0.001$ \\
MM & WM & $-3.7$ & $-7.57$ & $< 0.001$ & $-3.5$ & $-7.03$ & $< 0.001$ \\
MM & WW & $-12.5$ & $-18.05$ & $< 0.001$ & $-11.4$ & $-17.10$ & $< 0.001$ \\ \hline
MW & MM & $-0.4$ & $-1.92$ & $0.028$ & $-0.3$ & $-1.68$ & $0.046$ \\
MW & MW & $4.8$ & $4.05$ & $< 0.001$ & $4.2$ & $3.82$ & $< 0.001$ \\
MW & WM & $-0.0$ & $-0.02$ & $0.493$ & $-0.2$ & $-0.18$ & $0.429$ \\
MW & WW & $-1.9$ & $-1.35$ & $0.088$ & $-1.5$ & $-1.25$ & $0.105$ \\ \hline
WM & MM & $-0.7$ & $-4.40$ & $< 0.001$ & $-0.6$ & $-3.18$ & $< 0.001$ \\
WM & MW & $3.1$ & $3.64$ & $< 0.001$ & $2.1$ & $2.05$ & $0.020$ \\
WM & WM & $2.5$ & $3.35$ & $< 0.001$ & $2.2$ & $2.45$ & $0.007$ \\
WM & WW & $-0.7$ & $-0.60$ & $0.276$ & $-0.2$ & $-0.16$ & $0.438$ \\ \hline
WW & MM & $-3.1$ & $-10.56$ & $< 0.001$ & $-2.7$ & $-11.76$ & $< 0.001$ \\
WW & MW & $5.0$ & $3.48$ & $< 0.001$ & $3.8$ & $3.14$ & $< 0.001$ \\
WW & WM & $4.9$ & $4.05$ & $< 0.001$ & $4.2$ & $4.00$ & $< 0.001$ \\
WW & WW & $10.3$ & $8.59$ & $< 0.001$ & $9.6$ & $8.65$ & $< 0.001$ \\ \hline
Any & MM & $0.7$ & $9.18$ & $< 0.001$ & $0.7$ & $8.90$ & $< 0.001$ \\
Any & MW & $0.2$ & $0.40$ & $0.343$ & $-0.5$ & $-1.02$ & $0.153$ \\
Any & WM & $-1.9$ & $-4.59$ & $< 0.001$ & $-1.8$ & $-4.33$ & $< 0.001$ \\
Any & WW & $-7.2$ & $-12.13$ & $< 0.001$ & $-6.4$ & $-11.12$ & $< 0.001$ \\ \hline
 \end{tabular}
 \end{center}
\end{table}

\begin{table}[t]
   \caption{Comparison of the over/under-citation made by MM papers versus WW papers.}
   \label{table:s14}
 \begin{center}
\begin{tabular}{|c|c|c|c|c|} \hline
   Received by & Made by MM & Made by WW & $t$-statistic & $p$ value \rule[0pt]{0pt}{12pt} \\ \hline
MM & $1.24 \pm 0.18$ & $-2.27$ & $194.7$ & $< 0.001$ \\
MW & $-1.13 \pm 1.33$ & $5.62$ & $-50.6$ & $< 0.001$ \\
WM & $-2.40 \pm 1.16$ & $3.82$ & $-53.6$ & $< 0.001$ \\
WW & $-10.69 \pm 1.31$ & $7.78$ & $-141.1$ & $< 0.001$ \\
 \hline
 \end{tabular}
 \end{center}
\end{table}

\begin{table}[t]
\caption{Comparison of the over/under-citation made by conference papers versus journal papers for MM and WW papers.}
\label{table:s15}
\begin{center}
\begin{tabular}{|c|c|E|E|c|c|} \hline
Made by & Received by & Conference & Journal & $t$-statistic & $p$ value \rule[0pt]{0pt}{12pt} \\ \hline
\multirow{4}{*}{MM} & MM & $1.63$ & $1.10\ \pm\ 0.04$ & $-117.8$ & $< 0.001$ \\
 & MW & $-2.30$ & $-3.14\ \pm\ 0.35$ & $-24.1$ & $< 0.001$ \\
 & WM & $-6.06$ & $-3.84\ \pm\ 0.33$ & $66.9$ & $< 0.001$ \\
 & WW & $-18.30$ & $-9.52\ \pm\ 0.45$ & $194.2$ & $< 0.001$ \\ \hline
\multirow{4}{*}{MM} & MM & $-2.74$ & $-2.79\ \pm\ 0.29$ & $-1.6$ & $0.114$ \\
 & MW & $2.09$ & $7.63\ \pm\ 1.86$ & $29.8$ & $< 0.001$ \\
 & WM & $3.03$ & $-0.43\ \pm\ 1.49$ & $-23.2$ & $< 0.001$ \\
 & WW & $13.59$ & $11.68\ \pm\ 1.41$ & $-13.5$ & $< 0.001$ \\ \hline
\end{tabular}
\end{center}
\end{table}

\begin{table}[t]
\caption{Comparison of the over/under-citation made by papers with prominent authors versus those without for MM and WW papers.}
\label{table:s16}
\begin{center}
\begin{tabular}{|c|c|E|E|c|c|} \hline
Made by & Received by & With prominent authors & Without prominent authors & $t$-statistic & $p$ value \rule[0pt]{0pt}{12pt} \\ \hline
\multirow{4}{*}{MM} & MM & $1.56$ & $1.11\ \pm\ 0.10$ & $-44.2$ & $< 0.001$ \\
 & MW & $-3.86$ & $-2.04\ \pm\ 0.70$ & $26.0$ & $< 0.001$ \\
 & WM & $-5.10$ & $-4.36\ \pm\ 0.73$ & $10.3$ & $< 0.001$ \\
 & WW & $-19.17$ & $-11.55\ \pm\ 0.95$ & $79.9$ & $< 0.001$ \\ \hline
\multirow{4}{*}{WW} & MM & $-2.50$ & $-4.97\ \pm\ 0.63$ & $-39.4$ & $< 0.001$ \\
 & MW & $17.96$ & $8.61\ \pm\ 3.30$ & $-28.3$ & $< 0.001$ \\
 & WM & $4.97$ & $7.74\ \pm\ 3.30$ & $8.4$ & $< 0.001$ \\
 & WW & $-2.77$ & $18.66\ \pm\ 3.01$ & $71.2$ & $< 0.001$ \\ \hline
\end{tabular}
\end{center}
\end{table}

\begin{table}[t]
\caption{Comparison of the over/under-citation made by papers with high \(\text{MA}_{\text{or}}\) values versus those with low \(\text{MA}_{\text{or}}\) values for MM and WW papers.}
\label{table:s17}
\begin{center}
\begin{tabular}{|c|c|E|E|c|c|} \hline
Made by & Received by & Top half of $\text{MA}_{\text{or}}$ & Bottom half of $\text{MA}_{\text{or}}$ & $t$-statistic & $p$ value \rule[0pt]{0pt}{12pt} \\ \hline
\multirow{4}{*}{MM} & MM & $1.11$ & $1.66\ \pm\ 0.04$ & $122.6$ & $< 0.001$ \\
 & MW & $-0.97$ & $-5.13\ \pm\ 0.37$ & $-113.1$ & $< 0.001$ \\
 & WM & $-4.08$ & $-6.35\ \pm\ 0.34$ & $-67.3$ & $< 0.001$ \\
 & WW & $-10.87$ & $-15.87\ \pm\ 0.47$ & $-106.0$ & $< 0.001$ \\ \hline
\multirow{4}{*}{MM} & MM & $-3.56$ & $-1.87\ \pm\ 0.42$ & $40.6$ & $< 0.001$ \\
 & MW & $5.02$ & $5.52\ \pm\ 2.47$ & $2.0$ & $0.044$ \\
 & WM & $2.56$ & $-3.34\ \pm\ 1.86$ & $-31.7$ & $< 0.001$ \\
 & WW & $13.51$ & $16.10\ \pm\ 1.64$ & $15.7$ & $< 0.001$ \\ \hline
\end{tabular}
\end{center}
\end{table}

\begin{table}[t]
   \caption{Statistical significance of relationships among the paper's three categorical variables.}
   \label{table:s18}
 \begin{center}
\begin{tabular}{|B|D|D|A|A|} \hline
   Gender category & Characteristic $X$ & Characteristic $Y$ & $\chi^\textit{}2$-statistic & $p$ value \\ \hline
\multirow{6}{*}{MM} & Venue type & Participation of prominent authors & 4069.5 & $< 0.001$ \\ \cline{2-5}
 & Venue type & Local coauthorship network & 568.8 & $< 0.001$ \\ \cline{2-5}
 & Participation of prominent authors & Local coauthorship network & 924.2 & $< 0.001$ \\ \hline
\multirow{6}{*}{WW} & Venue type & Participation of prominent authors & 270.1 & $< 0.001$ \\ \cline{2-5}
 & Venue type & Local coauthorship network & 2.5 & $0.114$ \\ \cline{2-5}
 & Participation of prominent authors & Local coauthorship network & 320.7 & $< 0.001$ \\ \hline
 \end{tabular}
 \end{center}
\end{table}

\renewcommand{\refname}{Supplementary References}

\end{document}